\documentclass[prd,aps,twocolumn,superscriptaddress,floatfix,preprintnumbers,nofootinbib]{revtex4-1}
\usepackage{amsmath}
\usepackage{graphicx}
\usepackage{xcolor}
\usepackage{amsfonts}
\usepackage{amssymb,amsmath,bm}
\usepackage[citecolor=blue,pdfa=true,linktocpage=true,urlcolor=blue,colorlinks=true]{hyperref}
\usepackage[export]{adjustbox}
\usepackage{multirow}
\usepackage{physics}
\usepackage[capitalise]{cleveref}
\usepackage{soul}
\creflabelformat{equation}{#2\textup{#1}#3}
\crefname{section}{Sec.}{Secs.}

\newcommand{\aap}{{Astron.~Astrophys.}}

\newcommand{\mnras}{{Mon.~Not.~R.~Astron.~Soc.}}
\newcommand{\jcap}{JCAP}

\newcommand{\be}{\begin{equation}}
\newcommand{\ee}{\end{equation}}

\newcommand{\bea}{\begin{eqnarray}}
\newcommand{\eea}{\end{eqnarray}}
\newcommand{\bdm}{\begin{displaymath}}
\newcommand{\edm}{\end{displaymath}}

\def\fNL{f_{\mathrm{NL}}}

\newcommand{\bk}{\mathbf{k}}
\newcommand{\bn}{\mathbf{n}}

\def\bx{\bold{x}}

\def\bn{\bold{n}}
\def\br{\bold{r}}

\def\kpe{k_{\perp}}
\def\kpa{k_{||}}

\begin{document}

\title{Separating Angular and Radial Modes with Spherical-Fourier Bessel Power Spectrum on All Scales and Implications for Systematics Mitigation}

\author{Robin Y. Wen}
\email{ywen@caltech.edu}

\affiliation{California Institute of Technology, 1200 E. California Boulevard, Pasadena, CA 91125, USA}

\author{Henry S. Grasshorn Gebhardt}
\affiliation{California Institute of Technology, 1200 E. California Boulevard, Pasadena, CA 91125, USA}
\affiliation{Jet Propulsion Laboratory, California Institute of Technology, Pasadena, California 91109, USA}

\author{Chen Heinrich}
\affiliation{California Institute of Technology, 1200 E. California Boulevard, Pasadena, CA 91125, USA}

\author{Olivier Dor\'e}
\affiliation{California Institute of Technology, 1200 E. California Boulevard, Pasadena, CA 91125, USA}
\affiliation{Jet Propulsion Laboratory, California Institute of Technology, Pasadena, California 91109, USA}

\begin{abstract}
Current and upcoming large-scale structure surveys place stringent requirements on the mitigation of observational systematics in order to achieve their unprecedented constraining power. In this work, we investigate the potential use of the spherical Fourier-Bessel (SFB) power spectrum in controlling systematics, leveraging its capability of disentangling angular and radial scales. We first clarify how the discrete SFB basis describes radial scales via the index $n$ and demonstrate that the SFB power spectrum reduces to the clustering wedge $P(k,\mu)$ in the plane-parallel limit, enabling it to inherit results from past literature based on the clustering wedge. While the parallel and perpendicular Fourier mode $(k_{||}, k_\perp)$ decomposition underlying the wedge is only valid for surveys of small angular coverage with a well-defined global line-of-sight, the SFB basis provides a natural generalization that can be applied to the full sky. Crucially, the separation of angular and radial scales allows systematics to be localized in SFB space. In particular, systematics with broad and smooth radial distributions primarily concentrate in the $n=0$ modes corresponding to the largest radial scales. This localization behavior enables one to selectively remove only particular angular and radial modes contaminated by systematics. This is in contrast to standard 3D clustering analyses of wide-field surveys based on power spectrum multipoles, where systematic effects necessitate the removal of all modes below a given $k_{\rm min}$. Our findings advocate for adopting the SFB basis in 3D clustering analyses where systematics have become a limiting factor.
\end{abstract}
\maketitle

\section{Introduction}

Current and upcoming surveys such as DESI \cite{24DESI_II}, Euclid \cite{24Euclid_overview}, SPHEREx \cite{14SPHEREx}, LSST \cite{19LSST}, Roman \cite{20Eifler_Roman}, and SKA \cite{20SKA} will measure the large-scale structure (LSS) of the Universe over increasingly large volume, offering exciting prospects for constraining effects that manifest on large scales, such as local primordial non-Gaussianity (PNG)~\cite{01Komatsu_fNL,07Dalal_PNG,08Slosar_fnl,17dePutter_fnl1} and general relativistic (GR) effects~\cite{09Yoo_GR,10Yoo_GR,11ChallinorLPS,11Bonvin_GR,12JeongLPS}. The main challenge of extracting information from large scales is to understand and mitigate various observational systematics, which can confound the cosmological signal of interest. With the increasing survey volume enabling greater
statistical precision, the challenge of identifying and mitigating systematic contamination will become ever more important.

Two of the most prominent types of observational systematics in galaxy surveys are imaging systematics and redshift systematics. The imaging systematics can be caused by Galactic
foregrounds, such as dust extinction and stellar density, or varying imaging conditions, such as astrophysical seeing, photometric calibrations, and survey depth \cite{21eBOSS_fnl_catalog,20Kitandis_DESI_IC_systematics,23_DESI_LRG,22Chaussidon_DESILS_QSO}. Redshift systematics include all the potential effects associated with determining the redshift in either photometric \cite{21Cordero_redshift_uncertainty_marginal,22DES_BAO_3D} or spectroscopic surveys \cite{24Krolewski_spec}.

In the standard analysis using the power spectrum multipoles (PSM) $P_{L}(k)$ in Cartesian Fourier space, these systematics modulate the galaxy
power spectrum on large scales and often lead to an excess signal in the PSM (see, e.g.~\cite{23_DESI_LRG,24_DESI_Y1_PNG}), which can be misinterpreted as a signature of non-zero local PNG. However, the imaging and spectroscopic systematics fundamentally have different geometric characteristics compared to the large-scale cosmological signals. More specifically, the imaging systematics predominantly impact angular modes, while redshift systematics tend to be mostly radial in nature. On the other hand, a non-zero local PNG modulates both angular and radial modes. This difference in the angular-radial space between observational systematics and cosmological signals is lost in the Cartesian-based PSM, which mixes angular and radial fluctuations. As a result, the systematics and cosmological signals of interest can be easily misconstrued in the PSM.

While the clustering wedge $P(k,\mu)$ or its equivalent cylindrical power spectrum $P(\kpa,\kpe)$ have been used to disentangle angular and radial modes in Cartesian Fourier space, their validity is limited by the assumption of a fixed line-of-sight (LOS) across the survey volume. This approximation holds reasonably well on small scales in narrow-field surveys, but it breaks down in wide-field surveys where the LOS direction varies significantly. This breakdown is one of the key reasons why PSM—rather than the clustering wedge—have become the default statistics in Cartesian Fourier space, as they can still be reliably estimated over wide fields. Nevertheless, any Cartesian-based formalism ultimately fails to accurately characterize and disentangle large-scale angular and radial modes due to its inherent mismatch with the survey geometry.

This calls for a different basis that is spherical in nature rather than the frequently used Cartesian Fourier basis. One particular promising choice is the spherical Fourier-Bessel (SFB) basis--a combination of spherical harmonics for decomposing the angular dimensions, and spherical Bessel functions for the radial dimension. As the eigenfunctions of the Laplacian in the spherical coordinates, the SFB basis preserves the geometry of the curved sky \cite{18CastorinaWA,24PSM_SFB} and naturally incorporates effects on the light cone \cite{13Yoo_GR_SFB,24Semenzato_SFB_GR,24GR_SFB}. It was first introduced in the 90s \cite{91Binney_GaussianRF,93lahav_spherical} and used in the analysis of earlier galaxy~\cite{95Fisher_SFB,95Heavens_SFB,99Tadros_PSCz,04Percival_2DF_SFB} and weak lensing surveys~\cite{07Kitching_COMBO,14Kitching_SFB_CFHT}.

Despite the advantage of better matching the survey geometry, early SFB analyses tended to be plagued by numerical instabilities and computational complexity for both theory calculations and data estimation, which significantly limited its use in previous analyses. Recent work including Refs.~\cite{12Leistedt,19Samushia_SFB,22Khek_SFB_fast,21Gebhardt_SuperFab,23Gebhard_SFB_eBOSS} have made significant progress in addressing these numerical challenges in both estimator and theory. These advances have established the SFB basis as a feasible and practical alternative to the PSM for analyzing the large-scale clustering of current and upcoming LSS surveys. It is therefore prime time to investigate the advantages brought by the separation of angular and radial modes with the SFB basis, including the implications for systematics mitigation, which is the main goal of this work. 

To elucidate the separation of angular and radial modes\footnote{Throughout this work, we use the radial (line-of-sight) mode to refer to the oscillations at the radial direction in configuration space. This is not to be confused with the magnitude of the Fourier mode $k$, which is the radial length or the radial component of the Fourier mode $\bk$ in the Fourier space.} offered by the SFB basis, we will first provide some clarification to the interpretation of discrete SFB modes. We emphasize that the quantity $k$ in the SFB basis is not the radial mode in the configuration space but the total magnitude of the Cartesian Fourier mode $\bk$. The radial (LOS) mode is explicitly characterized by the index $n$ in the discrete SFB basis.

Using our understanding of radial modes, we study how the SFB power spectrum is related to the Cartesian Fourier PS. That is, we consider the plane-parallelization of the SFB PS. The plane-parallel approximation to (or the flat-sky limit of) the angular power spectrum, which describes projected clustering in 2D, has been well studied in the literature for both CMB \cite{00Hu_CMBLens,11Bernardeau_CMB_flat_sky} and LSS observations \cite{21Matthewson,20Gebhardt_non-linear,23Gao_flat_angular,24Gao_flat_angular}, while limited attempt has been made for the SFB power spectrum that characterizes 3D clustering. In this work, we will show the SFB power spectrum reduces to the clustering wedge $P(k,\mu)$ under the Cartesian Fourier basis in the limit of small angular scales or small radial scales.

In the context of both galaxy and line intensity mapping surveys, there have been numerous studies of observational systematics for narrow-field surveys using the Cartesian-based clustering wedge (e.g.~\cite{17BOSS_fourier_wedge,17Hand,17Pinol_wedge,14Liu_wedge1,16Cheng_interloper,19Gebhardt_interloper,16Lidz_interloper}). As the full-sky generalization of the clustering wedge, we expect the SFB power spectrum to inherit these past results and to be just as successful as the clustering wedge in localizing systematics in their respective modes while being applicable to wide-field surveys across all scales. 

The potential of the SFB basis for mitigating systematics in galaxy surveys has been recognized in previous work such as Refs.~\cite{18CastorinaWA,24PSM_SFB}. To our knowledge, Ref.~\cite{16SFB_IM} is the only work that has offered a detailed study of observational systematics in LSS surveys with the SFB basis. Compared to Ref.~\cite{16SFB_IM}, we consider galaxy surveys instead of only focusing on line intensity mapping surveys, and use the discretized SFB basis (given in \cref{eq:gnl_basis}) proposed in Refs.~\cite{19Samushia_SFB,21Gebhardt_SuperFab}, instead of the continuous SFB basis. The many advantages of the discrete SFB basis over the continuous SFB basis have been discussed in detail in Ref.~\cite{24GR_SFB}.

This paper is organized as follows: we first review the Cartesian Fourier and SFB formalism (Secs.~\ref{sec:Cartesian} and \ref{sec:SFB}) and discuss the properties of the SFB radial basis functions to clarify the nature of radial modes (Sec.~\ref{sec:radial}). In Sec.~\ref{sec:numerical-approx}, we will show how the SFB power spectrum can be numerically approximated by a plane-parallel Cartesian power spectrum, with analytical approximations discussed in Appendix~\ref{sec:analytical-approx}. In \cref{sec:additive-syst}, we illustrate the power of the SFB basis for localizing systematics in two cases: 1) angular or radial systematics, and 2) stellar contamination as a realistic foreground. We then discuss mitigation strategies and the connection between the clustering wedge and the SFB power spectrum in the context of systematics in \cref{sec:discussion}. We conclude in \cref{sec:conclusion}.

\section{Formalism}\label{sec:formalism}
We first review the Cartesian and the spherical Fourier-Bessel analysis of the 3D galaxy overdensity field. We follow Refs.~\cite{15Scoccimarro_FKP,17Hand,23WAGR} for the Cartesian basis in \cref{sec:Cartesian} and Refs.~\cite{19Samushia_SFB,21Gebhardt_SuperFab,23Gebhard_SFB_eBOSS} for the discrete SFB basis in \cref{sec:SFB}.

\subsection{Cartesian framework}\label{sec:Cartesian}
For an observed galaxy overdensity field $\delta(\bx)$ where $\bx$ is the comoving-space coordinate for the observed light cone, the Cartesian Fourier transform of the field is
\begin{align}
    \delta(\bk)=\int \dd^3\bx\; e^{-i\bk\cdot\bx}\delta(\bx)\,,
\end{align}
and one can form the Cartesian galaxy power spectrum
\begin{align}
    P(\bk,\bk')\equiv \langle \delta(\bk) \delta^*(\bk') \rangle\,,
    \label{eq:C-PS}
\end{align}
which is not diagonal due to redshift-space distortion, redshift evolution, and the geometry of observations \cite{96Zaroubi}. The Cartesian power spectrum only becomes diagonal when one considers the underlying matter distribution at a fixed redshift:
\begin{align}
    \langle \delta_{\rm m}(\bk,z) \delta_{\rm m}^*(\bk',z) \rangle = (2\pi)^3 \delta^{\rm D}(\bk-\bk')P_{\rm m}(k,z)\,.\label{eq:matter-PS}
\end{align}
Considering an infinite volume, the translational invariance and the statistical isotropy of the Universe only apply to the matter distribution at a fixed redshift . 

The matter power spectrum at a given redshift is the most natural quantity to obtain from theory, but it is not directly observable. The full 3D Cartesian power spectrum in \cref{eq:C-PS} is constructed from the observed field, but its exact computation is difficult due to the intrinsic mismatch between Cartesian coordinates and the spherical geometry of all the physical and observational effects taking place on the light cone. To more easily express the observed galaxy statistics in terms of the matter power spectrum in Cartesian Fourier space, one can introduce the local power spectrum \cite{15Scoccimarro_FKP,23WAGR}
\begin{align}
P^{\rm loc}(\bk,\bx_c)&\equiv\int \dd^3\bx_{12}\; e^{-i\bk\cdot\bx_{12}}\xi(\bx_c,\bx_{12})\,,
\label{eq:local-PS}
\end{align}
where $\bx_{12}\equiv\bx_1-\bx_2$ is the separation between the two galaxies, and $\xi(\bx_c,\bx_{12})\equiv \langle\delta(\bx_1)\delta(\bx_2)\rangle$ is the two-point correlation function parametrized in terms of the galaxy separation $\bx_{12}$ and some $\bx_c$ that defines the local LOS for the galaxy pair. The local LOS  can be simply taken as one of the galaxies in the pair, that is $\bx_c\equiv \bx_1$. 

Due to the azimuthal symmetry of the Universe, the local power spectrum only depends on 
\begin{align}
    P^{\rm loc}(\bk,\bx_c)=P^{\rm loc}(k,\mu,x_c)\,,
    \label{eq:Pkmu-definition}
\end{align}
where $\mu\equiv\hat{\bk}\cdot\hat{\bx}_c$ characterizes the angle between the Fourier vector and the local LOS. It is then possible to expand the local power spectrum in terms of Legendre polynomials and obtain the local power spectrum multipoles
\begin{align}
    P^{\rm loc}_{L}(k,x_c)=\frac{2L + 1}{2}\int^{1}_{-1}\dd\mu \,P^{\rm loc}(k,\mu,x_c) \mathcal{L}_L(\mu)\,.
    \label{eq:local-PSM}
\end{align}

On linear scales where the Finger-of-God (FoG) effect is small, the local power spectrum in the plane-parallel limit under the Newtonian redshift-space distortion (RSD) effect is described by the well-known Kaiser formula \cite{87Kaiser}
\begin{align}
P^{\rm loc}(k,\mu,x_c)&=\left(b_1+f\mu^2\right)^2P_m(k,x_c)\,,
\label{eq:P-RSD}
\end{align}
where $b_1$ is the linear galaxy bias, and $f$ is the linear growth rate. The Kaiser formula only applies in the plane-parallel limit when the galaxy separation remains small compared to the distance to the galaxy pair, that is $x_{12}/x_c\ll 1$. To go beyond the plane-parallel limit, one can obtain the exact expression of the local power spectrum and multipoles (with non-perturbative wide-angle corrections) through the correlation function \cite{22CatorinaGR-P,23Foglieni}. 

The Kaiser formula in \cref{eq:P-RSD} has been the main motivation for viewing the galaxy power spectrum as a function of the LOS Fourier-mode $k_{||}$ and the transverse Fourier mode $k_{\perp}$, which are defined as
\begin{align} k_{||}&\equiv\bk\cdot\hat{\bx}_{c}=\mu k\,,{\rm and}
    \label{eq:k-pa}\\
    k_{\perp}&\equiv \sqrt{k^2-k_{||}^2}\,.
    \label{eq:k-pp}
\end{align}
This decomposition of the Fourier mode is well-defined for any given LOS $\hat{\bx}_c$. For a redshift survey with small angular coverage, the survey geometry is roughly rectangular, and there is a well-defined global LOS for the whole field, which is known as the global plane-parallelization. In such a scenario, one can bin the 3D Fourier power spectrum $P(\bk)$ of the observed field along contours of constant $k_{\perp}$, obtaining the cylindrically binned power spectrum $P(k_{\perp},k_{||})$ that can be modeled by the Kaiser formula. 

However, for a survey with large angular coverage, there is no uniquely defined LOS for the whole observation, so both the transverse and LOS Fourier modes are no longer well-defined for the entire survey. To retain the Cartesian Fourier basis while still capturing the anisotropic clustering of the observed field due to RSD, it has become standard to use the Yamamoto estimator \cite{06Yamamoto} for measuring the power spectrum multipoles, where one of the galaxies is chosen as the local LOS for each galaxy pair (known as the local plane-parallel approximation). The PSM measured from the Yamamoto estimator can be modeled by the local PSM in \cref{eq:local-PSM} under the effective redshift approximation \cite{22CatorinaGR-P,24Cagliari_eBOSS_fnL_quasar}. The Yamamoto estimator has been widely used for galaxy surveys, since it captures the anisotropic clustering through decomposition with Legendre polynomials and enables efficient estimation with fast-Fourier transforms (FFT) \cite{15Scoccimarro_FKP,15Bianchi_FFT,17Hand}, with mature theoretical modeling on quasi-linear scales. However, by mixing angular and radial modes, the Cartesian Fourier transform used in the Yamamoto estimator becomes mismatched with the curved-sky geometry of wide-field survey, making it cumbersome for modeling the light-cone effects and mitigating observational systematics.

\subsection{SFB framework}\label{sec:SFB}

The Cartesian Fourier basis consists of the eigenfunctions of the Laplacian operator in Cartesian coordinates. In comparison, the discrete SFB basis is composed of eigenfunctions of the Laplacian in spherical coordinates, namely the product of the angular basis functions, spherical harmonics $Y^*_{\ell m}(\hat{\bn})$, and the radial basis functions $g_{\ell}(k_{n\ell}x)$, which were first introduced to cosmology by Ref.~\cite{19Samushia_SFB}. 

The radial basis functions are the linear combinations of the spherical Bessel functions of the first kind and second kind \cite{21Gebhardt_SuperFab}
\begin{align}
g_{n\ell}(x) = c_{n\ell} \, j_\ell(k_{n\ell}x) + d_{n\ell} y_\ell(k_{n\ell}x)\,,
\label{eq:gnl_basis}
\end{align}
where $k_{nl}$ denotes the total magnitude of each SFB mode, sharing the same meaning as $k$ in the Cartesian Fourier basis. The SFB modes are properly discretized under a finite comoving distance range $x_{\rm min}\leq x \leq x_{\rm max}$, with $n$ denoting the index for wavenumber $k_{n\ell}$ at each angular multipole $\ell$. The constants $c_{n\ell}$ and
$d_{n\ell}$ are chosen to satisfy the orthonormality relation
\begin{align}
\label{eq:gnl_orthonormality}
\int_{x_{\rm min}}^{x_{\rm max}}\dd{x} x^2g_{n\ell}(x)g_{n'\ell}(x)
&=
\delta^{\rm K}_{nn'}\,,
\end{align}
where $\delta^{\rm K}$ indicates the Kronecker delta. These constants are also subject to some boundary conditions, such as the potential or velocity boundary conditions described in Appendix D of Refs.~\cite{21Gebhardt_SuperFab} and \cite{23Gebhard_SFB_eBOSS} respectively. 

\begin{figure}[tbp]
\centerline{\includegraphics[width=0.92\hsize]{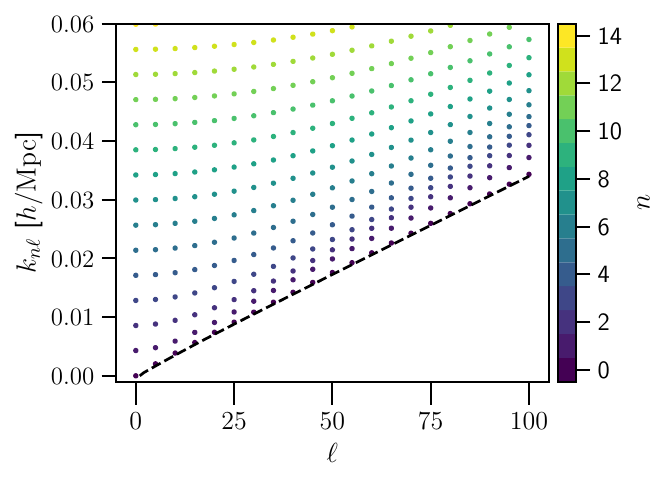}}
\caption{The values of the total Fourier $k_{n\ell}$ modes as a function of angular multipoles $\ell$ and radial indices $n$ for $z$ from $1.0$ to $1.5$ under the velocity boundary condition. The dashed black lines indicate the boundary formed by the purely angular $k_{0\ell}$ modes, which can be reasonably approximated by the linear equation $(\ell+\frac{1}{2})/x_{\rm max}$ as a function of $\ell$.}
\label{fig:knl}
\end{figure}

\begin{figure}[tbp]
\centerline{\includegraphics[width=0.85\hsize]{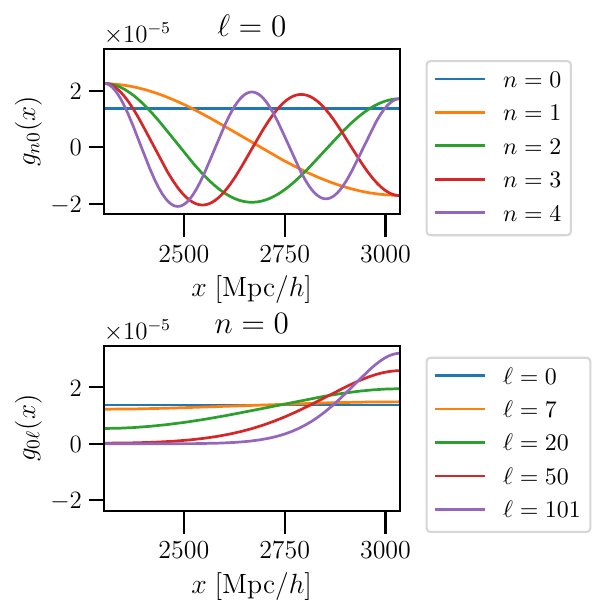}}
\caption{The SFB radial basis functions at $\ell=0$ and $n=0$ respectively for $z$ from $1.0$ to $1.5$ under the velocity boundary condition. For the purely radial modes with $\ell=0$, the SFB radial functions reduce to exactly sinusoidal waves with their amplitudes modulated by $1/x$ (see \cref{sec:analytical-radial}). For purely angular modes with $n=0$, the radial basis functions are all positive and do not cross zero.}
\label{fig:gnl_zero}
\end{figure}

\begin{figure}[tbp]
\centerline{\includegraphics[width=\hsize]{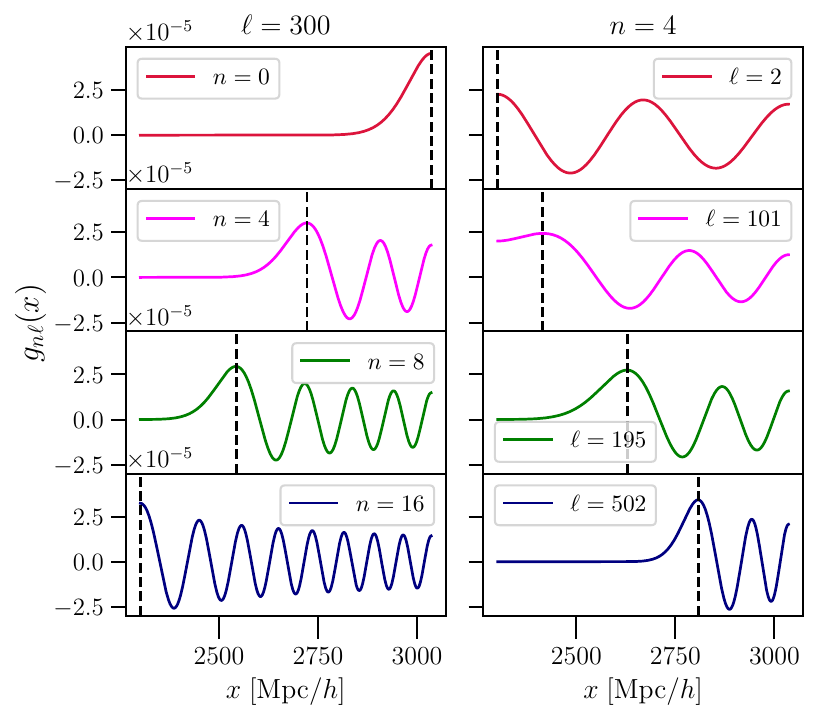}}
\caption{The SFB radial basis functions for different combinations of $\ell$ and $n$. Not all radial oscillations cover the entire redshift range where the SFB basis is defined, and the dashed black lines indicate the transition distance $x^{\rm t}_{n\ell}$ (defined in \cref{eq:x-transition}) where the radial oscillations actually start for each $k_{n\ell}$ mode.}
\label{fig:gnl_pattern}
\end{figure}

In this work, we choose the velocity boundary condition as the default, which we justify in \cref{sec:velocity-over-potential}. The velocity boundary condition (also known as the Neumann boundary condition) requires \cite{95Fisher_SFB,23Gebhard_SFB_eBOSS}
\begin{align}
\label{eq:velocity-boundary-def}
\begin{split}
    g_{\ell}'(kx_{\rm min})&=0\,,{\rm and}\\
    g_{\ell}'(kx_{\rm max})&=0\,,
\end{split}
\end{align}
where the derivatives of the radial basis functions vanish at the boundaries. The total Fourier magnitudes $k_{n\ell}$ and the radial basis functions $g_{n\ell}(x)$ under the velocity boundary condition for a redshift bin $z=1.0-1.5$ (corresponding to the comoving distance $x=2301-3036\,{\rm Mpc}/h$) are plotted in \cref{fig:knl,fig:gnl_zero,fig:gnl_pattern}. We will discuss the properties of these radial functions and the meaning of the discretized index $n$ in detail in \cref{sec:radial}.

The discrete SFB decomposition of the overdensity field $\delta(\bx)$ and the inverse transformation then become 
\begin{align}
\delta_{n \ell m}
&= \int \dd^3\bx\,g_{n\ell}(x)\,Y^*_{\ell m}(\hat{\bn})\delta(\bx)\,,{\rm and}
\label{eq:sfb_discrete_fourier_pair_b}\\
\delta(\bx)
&= \sum_{n\ell m} g_{n\ell}(x)\,Y_{\ell m}(\hat{\bn})\,\delta_{n\ell m}\,,
\label{eq:sfb_discrete_fourier_pair_a}
\end{align} 
where the integral of \cref{eq:sfb_discrete_fourier_pair_b} goes over the finite volume within $x_{\rm min}\leq x \leq x_{\rm max}$ across the full solid angle.

We can then form the discrete SFB power spectrum through
\begin{align}
   \langle\delta_{n_1 \ell_1 m_1}\delta_{n_2 \ell_2 m_2}^{*}\rangle = C_{\ell_1 n_1n_2}\delta_{\ell_1\ell_2}^{\rm K}\delta_{m_1m_2}^{\rm K}
   \label{eq:SFB-discrete-PS}\,.
\end{align}
The translational invariance is broken in the presence of redshift evolution and RSD, while the rotational invariance is still preserved, leading to the angular mode $\ell$ and two wavenumbers $n_1,n_2$ in \cref{eq:SFB-discrete-PS} as a complete decomposition of the two-point statistics for the overdensity field. The exact computation of the theoretical SFB power spectrum (PS) is computationally challenging even under linear theory due to the nested integration over four oscillatory Bessel functions. This problem has been partially addressed in Ref.~\cite{23Gebhard_SFB_eBOSS} through introducing the Iso-qr integration\footnote{Iso-qr integration stands for 1D spherical Bessel integration along constant $q$-$r$ lines.} method to compute the SFB PS under the linear Newtonian RSD effects, achieving evaluation of the SFB PS on the order of seconds on a single CPU. 

On the data side, the SFB power spectrum can be estimated through the pseudo-$C_{\ell}$ method  \cite{21Gebhardt_SuperFab}
\begin{align}
\widehat{C}_{\ell n_1n_2}\equiv\frac{1}{2\ell+1}\sum_{m}\delta_{n_1\ell m }\delta^*_{n_2\ell m }\,,\label{eq:SFB-estimator}
\end{align}
which is implemented in the public code \href{https://github.com/hsgg/SphericalFourierBesselDecompositions.jl}{\texttt{SuperFaB}}. Here the sum over the azimuthal index $m$ is analogous to the averaging of the Fourier modes in the plane perpendicular to the LOS in the global plane-parallel estimator for the cylindrical power spectrum.

\section{Radial Modes in SFB Basis}\label{sec:radial}
Exploiting the azimuthal symmetry and averaging over the index $m$ (\cref{eq:SFB-discrete-PS,eq:SFB-estimator}), we need two indices $\ell$ and $n$ to characterize the SFB modes in the power spectrum: $\ell$ is the angular multipole of spherical harmonics, while $n$ is produced due to the discretization of the total Fourier modes under the orthonormality relations and certain boundary conditions in the radial direction. The discretization of $k$ is different for each angular multipole $\ell$ due to the different behavior of the spherical Bessel functions of different order, so both $\ell$ and $n$ are needed to specify the discretized total Fourier modes, the patterns of which are shown in \cref{fig:knl}. 

Since $\ell$ is the angular multipole in spherical harmonics, it has the familiar meaning of an angular mode on the curved sky, the same meaning shared by the widely-used angular power spectrum. In comparison, the meaning of $n$ is not directly clear besides discretizing the Fourier magnitudes, and its exact meaning has not been extensively discussed in previous literature. Building upon Refs.~\cite{19Samushia_SFB,21Gebhardt_SuperFab,23Gebhard_SFB_eBOSS}, our goal here is to clarify and firmly establish the index $n$ for radial modes.

For a fixed angular multipole, we plot $g_{n\ell}(x)$ under different values of $n$ in the first panels of \cref{fig:gnl_zero,fig:gnl_pattern} for $\ell=0$ and $\ell=300$ respectively. We can see $n$ represents the number of radial oscillations in the SFB radial basis functions, and an increasing value of $n$ captures information at smaller scales in the radial direction. More specifically, $n$ is the number of zero-crossings, and, thus, represents the number of half-cycles for a given mode, provided the first mode starts at $n=0$. For $\ell=0$, the radial basis functions closely resemble the 1D Cartesian Fourier basis, which we confirm analytically in \cref{sec:analytical-radial}. These $k_{n0}$ modes (with angular monopole $\ell=0$) are essentially purely radial modes since the SFB basis functions do not possess any angular oscillation.

We emphasize that we have chosen to label the first and the largest radial mode as $n=0$ instead of $n=1$ as done in previous work\footnote{We first switched to the $n=0$ convention in Ref.~\cite{24GR_SFB}, and we here give the detailed justification for this choice.} \cite{21Gebhardt_SuperFab,23Gebhard_SFB_eBOSS}. We show the $n=0$ modes in the black dashed line of \cref{fig:knl}, and these modes can be roughly approximated by the line
\begin{align}
   k_{0\ell}\approx \frac{\ell+\frac{1}{2}}{x_{\rm max}},
\end{align}
which indicates the $n=0$ modes to be purely angular modes. These $k_{0\ell}$ modes also provide the lower bound for the magnitudes of all the SFB modes at a fixed $\ell$, since the presence of any radial oscillations on top of purely angular modes will further increase the value of the Fourier modes $k$.

We plot the radial basis functions for the $n=0$ modes in the lower panel of \cref{fig:gnl_zero}, and we observe that these functions do not possess any oscillation in the radial direction. For lower $\ell$ corresponding to larger angular scales, the basis function is essentially flat without any sign of oscillation. For higher $\ell$, $g_{0\ell}(x)$ increases as $x$ increases, which resembles the half cycle of a sinusoidal oscillation. However, we emphasize that such a visual appearance does not actually correspond to any radial oscillation, since the basis functions have no zeros. Such an increase is simply the smooth transition of the radial basis functions between the region where the radial mode vanishes and the plateau where the $n=0$ modes reside. Therefore, the first radial modes across all angular multipoles possess no radial oscillations, for which the $n=0$ label provides a better description. 

We next examine the radial basis functions under higher angular and radial modes, which are plotted in \cref{fig:gnl_pattern}. For any SFB mode, there exists a distance scale, that we denote as $x_{nl}^{\rm t}$, at which the behavior of the radial basis function changes: $g_{n\ell}(x)$ vanishes in the region $x<x^{\rm t}_{n\ell}$, while it oscillates for $x>x^{\rm t}_{n\ell}$. We can see in \cref{fig:gnl_pattern} that $g_{n\ell}(x)$ naturally interpolates between these two distinct regions separated by $x^{\rm t}_{n\ell}$. The radial basis functions vanish at the lower radius because our SFB basis is the eigenfunctions of the Laplacian over a spherical shell, which has more volume at a larger radius and therefore more modes at a larger radius. To decrease the effective number of modes at lower radius, the radial basis functions of certain modes have to vanish at lower redshift and only be present at the higher redshift portion of the redshift bin. 

We formally define the transition distance $x^{\rm t}_{n\ell}$ to be the actual start of the radial oscillations for a particular $k_{n\ell}$ mode, and for any $n>0$ modes\footnote{There is no radial oscillation for the $n=0$ modes, so the definition of the transition scale is a bit arbitrary for these modes. We use $x^{\rm t}_{0\ell}\equiv {\rm min}_x\{g_{0\ell}(x)>\frac{1}{4}g_{0\ell}(x_{\rm max})\}$ to represent the transition scale, which is found to provide reasonable approximations to the SFB power spectrum under the plane-parallel limit in \cref{sec:numerical-approx}.}, the transition distance can be mathematically characterized by
\begin{align}
    x^{\rm t}_{n\ell}\equiv {\rm min}_x\{g_{n\ell}'(x)<0\}\,,
    \label{eq:x-transition}
\end{align}
the smallest distance where the derivative of the radial basis function becomes negative. The transition distance can be determined numerically, and we show the transition distance with the black dashed lines in \cref{fig:gnl_pattern}. 

As seen in the first column of \cref{fig:gnl_pattern}, the transition distance $x^{\rm t}_{n\ell}$ decreases as $n$ increases for a fixed angular mode $\ell$. The radial oscillations start at the high redshift portion of the redshift bin and gradually extend to the low redshift portion to cover the entire redshift bin as $n$ increases. For a fixed radial mode $n$, $x^{\rm t}_{n\ell}$ increases as $\ell$ increases (second column of \cref{fig:gnl_pattern}), indicating an increasing impact of angular modes on the behavior of the radial basis function. 

Based on the transition distance of each SFB mode, we can identify the analog to the LOS Fourier modes (defined in \cref{eq:k-pa}) present in each SFB mode as\footnote{The definition for the LOS Fourier mode in \cref{eq:k-LOS} is also motivated by the analytical approximations to the radial basis functions discussed in \cref{sec:analytical-radial}}
\begin{align}
    k_{||,n\ell}\equiv\frac{n\pi}{x_{\rm max}-x^{\rm t}_{n\ell}}\,,
    \label{eq:k-LOS}
\end{align}
which is determined by the ratio between the number of half-cycles of the radial oscillations and the distance of the region where the radial modes reside. \Cref{eq:k-LOS} explicitly gives the LOS Fourier mode in each SFB mode according to the index $n$. We can then define the transverse (perpendicular or angular) Fourier modes $k_{\perp,n\ell}$ in each SFB mode following \cref{eq:k-pp}. Note that our transverse Fourier modes definition depends on the LOS Fourier modes (\cref{eq:k-LOS}) based on the radial index $n$. Alternatively, it is also possible to directly approximate the transverse Fourier modes according to the angular multipoles $\ell$, which we discuss in \cref{sec:angular-approx}.

It has been a common misconception to perceive $k_{n\ell}$ as the radial (LOS) Fourier mode in the SFB basis. We emphasize that $k_{n\ell}$ is the magnitude of the total Fourier mode that includes both angular and radial contributions. Corresponding to the number of half oscillations in the radial basis functions, the index $n$ represents the properly discretized radial modes and determines the radial Fourier value present in each SFB mode through \cref{eq:k-LOS}. Therefore, the separation between angular and radial scales is completely achieved through the two indices $\ell$ and $n$ in the discrete SFB basis.

\section{Plane-Parallel Approximations of SFB Power Spectrum}\label{sec:numerical-approx}

\begin{figure*}[tbp]
\centerline{\includegraphics[width=0.87\hsize]{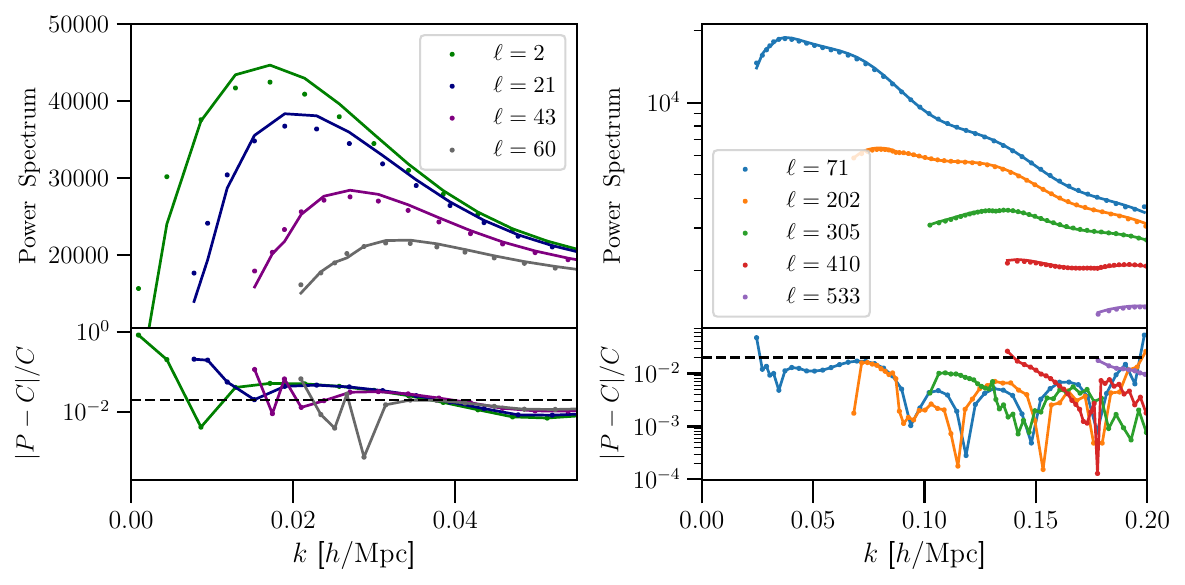}}
\caption{Comparison between the exact SFB PS and the corresponding plane-parallel $P(k_{n\ell},\mu_{n\ell})$ under the linear-order Newtonian RSD for $z=1.0-1.5$. The dots indicate the exact SFB power spectrum, while the solid lines indicate the Cartesian power spectrum approximations given by \cref{eq:pp-SFB}. The bottom panel shows the absolute fractional difference between the two spectra, and the black dashed lines indicate $2\%$ relative error. The left plot shows small angular multipoles in the wide-angle regime, while the right plot shows large angular multipoles in the plane-parallel regime. We see that the approximation scheme of \cref{eq:pp-SFB} achieves percent-level accuracy at higher angular multipoles across the $k$ range.}
\label{fig:SFB_compare_Pkmu}
\end{figure*}

Using our understanding of the radial modes in the SFB basis, we now focus on the numerical approximation of the SFB power spectrum under the plane-parallel limit. We will give the analytical justification for our results in \cref{sec:analytical-approx}. To our knowledge, the plane-parallelization of the SFB PS has only been considered in Ref.~\cite{19Chakraborty_21cm_wl}, but they use the continuous SFB basis while we study the discrete case, and our final expressions are more succinct by directly connecting back to the 3D Cartesian power spectrum. 

Using the LOS Fourier mode defined in \cref{eq:k-LOS}, we approximate the SFB power spectrum with the local Cartesian clustering wedge $P^{\rm loc}(k,\mu,x_c)$ (defined in \cref{eq:Pkmu-definition}) as 
\begin{align}
     C_{\ell nn}^{\rm approx.}\equiv P\left(k=k_{n\ell},\mu=\frac{k_{||,n\ell}}{k_{n\ell}},x_c=x_{{\rm eff},n\ell}\right)\,.
    \label{eq:pp-SFB}
\end{align}
Here the effective distance $x_{\rm eff}$ (similar to the effective redshift approximations used in calculating PSM) is defined for each $k_{n\ell}$ mode
\begin{align}
    x_{{\rm eff},n\ell}=\frac{\int^{x_{\rm max}}_{x^{\rm t}_{n\ell}}\dd x\,x^2R(x)^2}{\int^{x_{\rm max}}_{x^{\rm t}_{n\ell}}\dd x\,xR(x)}\,,
    \label{eq:eff-xnl}
\end{align}
where $R(x)$ is the radial selection function. The effective distance is only calculated over the range $[x^{\rm t}_{n\ell},x_{\rm max}]$ where the SFB mode resides. The average in the above is calculated over the measure $x\,dx$ (instead of the usual measure $x^2\,dx$ in the radial direction), which accounts for the $1/x$ amplitude that modulates the sinusoidal wave as seen in the approximation of the radial basis function given in \cref{eq:gp-kx}. Our approximation scheme directly applies to SFB PS covering the full sky with no angular mask. We will also assume a uniform radial selection in comoving space. In the case with realistic angular masking and radial selection, one can first approximate the unwindowed SFB PS using \cref{eq:pp-SFB} and then apply window convolution following Refs.~\cite{21Gebhardt_SuperFab,23Gebhard_SFB_eBOSS} to the approximated SFB PS.

Under the plane-parallel limit, Cartesian clustering wedge can be evaluated using the Kaiser formula at the effective distance defined in \cref{eq:eff-xnl}. To assess the accuracy of the above approximation, we exactly compute the SFB power spectrum under linear Newtonian RSD following Refs.~\cite{23Gebhard_SFB_eBOSS,24GR_SFB}. We assume a best-fit Planck 2018 $\Lambda$CDM cosmology \cite{18Planck_Parameter} and a constant linear galaxy bias $b_1=1.5$ for the galaxy power spectra computed in this work.

We note that our approximation only applies to the diagonal component of the SFB power spectrum (where $n_1=n_2$ such that $k_{n_1\ell}=k_{n_2\ell}$), which dominates over the off-diagonal components ($n_1\neq n_2$), as we show in \cref{fig:SFB_foreground_square,fig:SFB_foreground_line}. The effective redshift approximation ignores the redshift evolution within the redshift bin, while the plane-parallel approximation treats the two differing LOS of the two galaxies in a pair as the same, so the above approximation does not produce any off-diagonal components\footnote{See \cref{sec:analytical-PS} for analytical justification for the lack of off-diagonal components.}.

We compare the numerical results between the exact, diagonal SFB power spectrum $C_{\ell nn}$ and the approximated results $C_{\ell nn}^{\rm approx.}$ in \cref{fig:SFB_compare_Pkmu}. For each fixed angular multipole $\ell$, larger $k$ values towards the right of the plot represent larger $n$ values and smaller radial modes for the SFB PS. In the right panel, we show the comparison at large angular multipoles representing small angular scales. Here we see that both PS agree at the percent level across all $k$ scales for each $\ell$. This numerically confirms SFB power spectrum behaves as Cartesian clustering wedge under the plane-parallel limit at small angular scales. In the left panel, however, we show the comparison at small angular multipoles, where the SFB power spectrum differs significantly from its plane-parallel approximation. This is expected since the SFB basis correctly captures the spherical geometry for galaxies with large angular separation, while the plane-parallel assumption of the Kaiser formula breaks down \cite{18CastorinaWA,24Benabou_WA}.

In particular, the approximation is the most inaccurate for the first few $n$ modes at each angular $\ell$ where the transverse Fourier mode $k_{\perp,n\ell}$ has a higher value than the LOS Fourier mode $k_{||,n\ell}$. When the transverse modes contribute more to the total Fourier modes at a fixed angular separation, the wide-angle effects become more prominent, since the modes are effectively more angular than radial. The accuracy of the plane-parallel approximation significantly improves as $n$ increases and the LOS Fourier modes contribute more to the total Fourier modes, that is $k_{||,n\ell}\to k_{n\ell}$ as $n\to \infty$. 

The comparison of the two power spectra shown in \cref{fig:SFB_compare_Pkmu} demonstrates the advantage of the SFB basis at large angular and radial scales (lower $\ell$ and $n$ indices) over the Cartesian basis, while the SFB power spectrum reduces to the familiar Cartesian power spectrum $P(k,\mu)$ at small angular or small radial scales. Therefore, the exact computation of the SFB PS is only necessary at these large angular and large radial scales, where the linear-order calculation of the SFB PS is sufficient and has been completely developed. The regime where the plane-parallel approximation breaks down exactly indicates where an SFB-based analysis becomes more natural and beneficial compared to a Cartesian Fourier analysis.

The similarity between the Cartesian and SFB PS (albeit with the continuous SFB basis) is first noticed for astrophysical (angular) foregrounds in the context of line intensity mapping surveys~\cite{16SFB_IM}. We have now demonstrated this similarity for the power spectra of cosmological signals and switched to the discrete SFB basis. Our approximation scheme  firmly establishes the standard Cartesian Fourier basis as the plane-parallel limit of the SFB basis, thereby enabling us to transfer results and intuition from the well-established Cartesian framework to the less familiar SFB formalism. In particular, \cref{eq:pp-SFB} helps to interpret each mode present in the SFB PS $C_{\ell nn}$ through the familiar Cartesian power spectrum $P(k,\mu)$.

In addition, \cref{eq:pp-SFB} can be used to approximate the non-linear corrections to the SFB power spectrum, for which the one-loop correction at the quasi-linear scale is still missing in the literature, thereby limiting any SFB analysis to the strictly linear regime ($k\lesssim 0.05\,h/{\rm Mpc}$) \cite{23Gebhard_SFB_eBOSS}. At smaller scales where the non-linear contribution becomes important, one can calculate the Cartesian power spectrum $P(k,\mu)$, of which the one-loop evaluation under the effective field theory has become mature \cite{20Chudaykin_classpt,21Damico_EFT,24Linde_class_oneloop,21Chen_LPT}, and then map to SFB space following \cref{eq:pp-SFB}. Such a procedure will potentially achieve one-loop evaluations of the SFB PS at the percent-level accuracy at smaller scales. This will then enable an SFB-based analysis across the linear and quasi-linear scales. Although primarily intended for the analysis of clustering at large scales where the survey geometry becomes spherical, there are potential motivations to also adopt the SFB basis at smaller scales in the context of systematics, which we will discuss in \cref{sec:compare-wedge}. We defer the implementation and validation of such an approximation scheme to the 1-loop evaluation of the SFB power spectrum to future work.

\section{Observational Systematics}\label{sec:additive-syst}
We have so far clarified the separation of angular and radial scales into the two indices $\ell$ and $n$ in the SFB power spectrum and shown that it reduces to the Cartesian clustering wedge $P(k,\mu)$ in the plane-parallel limit. Both of these results motivate our study of observational systematics in the SFB basis in this section: the angular-radial separation offered by the SFB basis allows for isolating the observational systematics into particular SFB modes, while the connection to the Cartesian power spectrum allows us to transfer existing results of examining systematics through the clustering wedge $P(k,\mu)$ into the SFB basis.  

Here we first introduce our systematics model and explain why we focus only on additive systematics effects in the density contrast (\cref{sec:systematic-model}). We then study the SFB PS of angular (\cref{sec:angular-foreground}) or radial systematics (\cref{sec:radial-foreground}), and examine stellar contamination as a realistic example in \cref{sec:star}. 

\subsection{Systematics model}\label{sec:systematic-model}

We first clearly define our notation describing how observational systematics impact the observed galaxy overdensity field. We assume the systematics-contaminated galaxy number density observed by us is related to the underlying galaxy number density as
\begin{align}
    N^{\rm obs}(\bx)=N(\bx)(1+M(\bx))+A(\bx)\label{eq:sys}\,,
\end{align}
where $A(\bx)$ represents additive components from spurious objects, and $M(\bx)$ represents the effects of multiplicative selection on the galaxy number density. Here we assume both these functions are not correlated with the underlying galaxy density field.

In galaxy surveys, stellar contamination mostly produces an additive effect by introducing spurious objects into the galaxy catalog, which at the first order only depends on stellar densities. The effects of most other systematics such as dust extinction and survey depths modulate the observed galaxy densities, so these systematics can be modeled as multiplicative. These multiplicative effects on the galaxy number densities are commonly encapsulated in what is known as the selection function. For line intensity mapping surveys such as the ones using the 21 cm hyperfine transition of neutral hydrogen, astrophysical foregrounds such as the synchrotron emission from the Milky Way and extragalactic continuum sources are also additive and are several orders of magnitude brighter than the cosmological HI signal (hyperfine transition line of the neutral hydrogen). 

In terms of the underlying galaxy overdensity contrast defined as $N(\bx)=\overline{N}(1+\delta(\bx))$, our systematics model of \cref{eq:sys} produces
\begin{align}
    &\delta^{\rm obs}=\frac{N^{\rm obs}(x)-\overline{N}^{\rm obs}}{\overline{N}^{\rm obs}}\nonumber\\
    &=\frac{\overline{N}(1+\delta(\bx))(1+M(\bx))+A(\bx)}{\overline{N}^{\rm obs}}-1\nonumber\\
    &=\frac{\overline{N}}{\overline{N}^{\rm obs}}\delta(x)(1+M(\bx))\nonumber\\
    &\quad+\frac{\overline{N}}{\overline{N}^{\rm obs}}M(\bx)+\frac{1}{\overline{N}^{\rm obs}}A(\bx)-\left(1-\frac{\overline{N}}{\overline{N}^{\rm obs}}\right)\label{eq:systematics-density-contrast}\,,
\end{align}
which is similar to the expression in Ref.~\cite{24Berlfein_systematics}.

One can see that an additive component to the galaxy number density remains an additive component in the number density contrast. In comparison, a multiplicative effect on number density generates both additive and multiplicative components on the density contrast. With $\delta(\bx)\ll 1$ at the large scales, the additive effect due to the multiplicative selection $M(\bx)$ on the density contrast is orders of magnitude larger than the corresponding multiplicative effect. Therefore, our systematics model produces a predominantly additive effect in the density contrast. 

In this section, we focus on the additive effects on the density contrast, examining them through the SFB basis. The multiplicative effects on the density contrast in \cref{eq:systematics-density-contrast} are similar to the effect of a window function, and can be modeled in the SFB basis in the same way as for a window convolution following Refs.~\cite{21Gebhardt_SuperFab,23Gebhard_SFB_eBOSS}.

\begin{figure*}[tbp]
\centerline{\includegraphics[width=0.8\hsize]{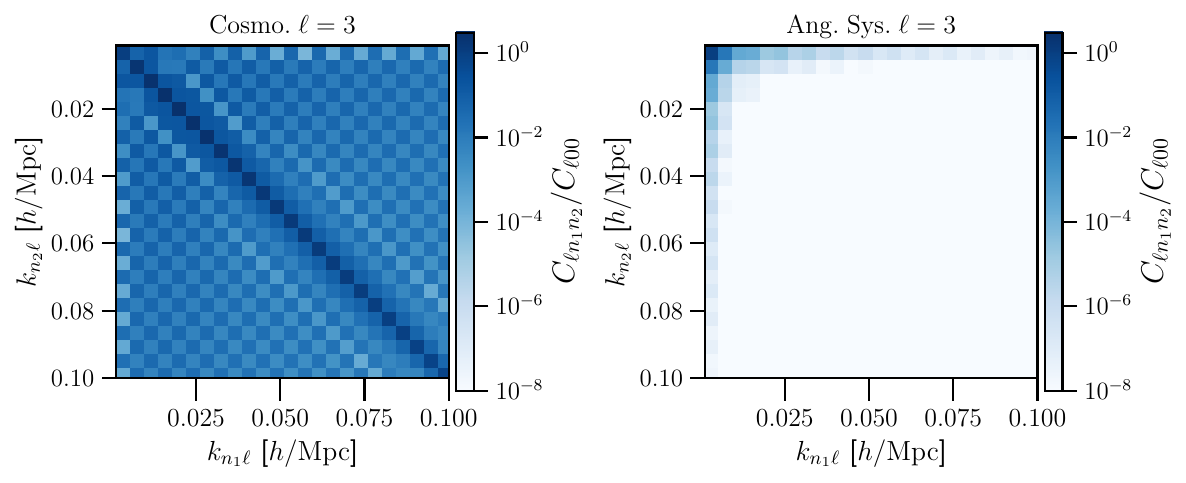}}
\caption{Comparison between the cosmology (abbreviated as ``cosmo.'', calculated assuming Planck 2018 $\Lambda$CDM cosmology with $f_{\rm NL}=0$) and angular systematics signals (abbreviated as ``ang. sys.", assuming uniform radial distributions) for the SFB power spectra, normalized by the $n_1=n_2=0$ modes at $\ell=3$ for $z=1.0-1.5$. The relative strengths of the signals with numerical values smaller than $10^{-8}$ are not shown.   
}
\label{fig:SFB_foreground_square}
\end{figure*}

\begin{figure*}[tbp]
\centerline{\includegraphics[width=0.775\hsize]{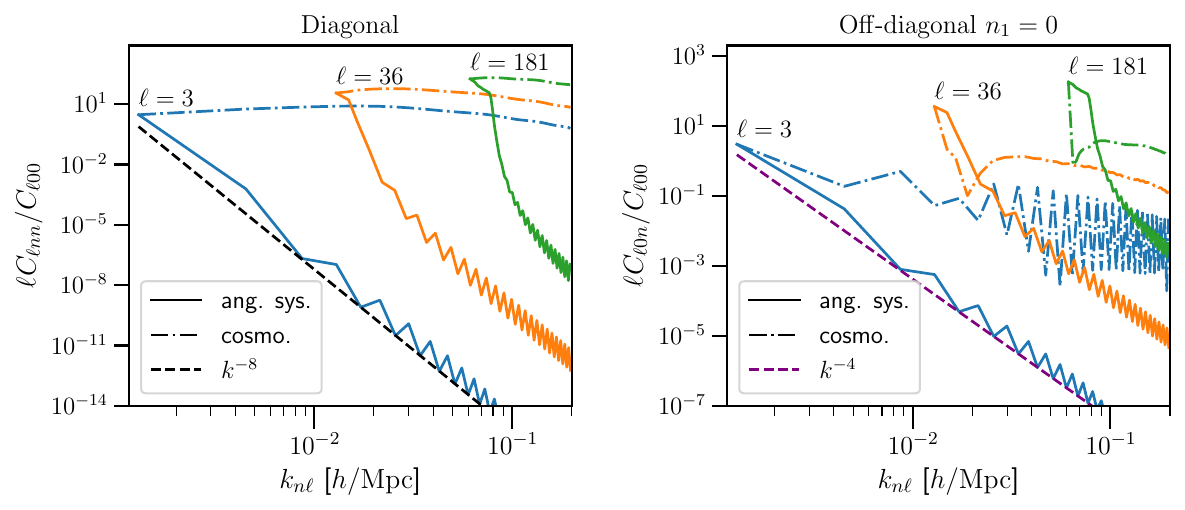}}
\caption{Comparison between the cosmology and angular systematic signals (with uniform radial distributions) for the diagonal and off-diagonal components of the SFB power spectrum, normalized by the $n_1=n_2=0$ modes, at different angular multipoles. We multiply the relative signal by $\ell$ to visually distinguish the curves corresponding to different angular multipoles. In both panels, the solid lines represent the systematics, while the dashed-dot lines represent the cosmological signal. The dashed lines show the reference lines indicating the scaling of the angular systematics with respect to $k$. The oscillations of the solid lines for angular systematics are due to the parity dependence, that is the even (odd) $n$ values share similar trends for a fixed $\ell$.
}
\label{fig:SFB_foreground_line}
\end{figure*}

\subsection{Angular systematics}\label{sec:angular-foreground}

Systematic effect can be classified as angular, radial, or as a combination of both. While this classification is a useful conceptual and modeling tool, it is important to recognize that the reality is more nuanced. Most systematics are neither purely angular nor purely radial—their influence on observed galaxy number densities is inherently three-dimensional, arising from their chromatic nature and the complex interplay between the galaxy spectral energy distribution (SED) and the survey’s color and magnitude selection. However, due to their origin from distinct physical or observational processes, systematics tend to affect angular and radial scales differently. This contrasts with primordial cosmological perturbations, which are statistically isotropic and do not distinguish between angular and radial modes. For this reason, it is useful to differentiate between systematics that primarily impact angular scales and those that predominantly affect radial scales, even if the classification is not perfectly clean. We refer to these, respectively, as angular or radial systematics.

We first consider angular systematics that produce additive effects on the density contrast $\delta$. Examples of these are fiber collision \cite{17Hahn_fiber,24Pinon_DESI_fiber_angle_cut} and the effects of imaging systematics such as stellar contamination and varying imaging conditions in galaxy surveys, which have been treated as predominantly angular so far \cite{24_DESI_Y1_PNG,23_DESI_LRG}, as well as astrophysical foregrounds in intensity mapping surveys. Our general strategy is similar to Section~5 of Ref.~\cite{16SFB_IM}, which studied the angular foreground in the SFB basis for line intensity mapping. Compared to Ref.~\cite{16SFB_IM}, we use the discrete SFB basis (given in \cref{eq:gnl_basis}) instead of the continuous SFB basis that only depends on the spherical Bessel function of the first kind. The discrete basis matches with the estimators that can be used on data and, as we will see, help to localize systematics into particular modes.

An angular systematic effect only depends on the spherical harmonic mode of the systematic $S_{\ell m}$, which is independent of the radial modes
\begin{align}
    S_{n \ell m}&=d_{n\ell}S_{\ell m},
\end{align}
where $d_{n\ell}$ are the radial coefficients of the unit line under the SFB basis
\begin{align}
    d_{n\ell}=\int_{x_{\rm min}}^{x_{\rm max}}\dd x\,x^2g_{n\ell}(x)\,,
    \label{eq:dnl}
\end{align}
and we have used the relationship between spherical harmonic and SFB modes in \cref{eq:SFB-TSH}.
Then the SFB PS obtained from \cref{eq:SFB-estimator} for the angular systematics is 
\begin{align}
    C^{S}_{\ell n_1n_2}&\equiv \frac{1}{2\ell+1}\sum_{m} S_{n_1 \ell m} S^*_{n_2 \ell m}=d_{n_1\ell}d_{n_2\ell}C^{S}_{\ell}\,,
    \label{eq:SFB-PS-angular-foregorund}
\end{align}
where $C^{S}_{\ell}\equiv \frac{1}{2\ell+1}\sum_{m} S_{\ell m} S^{*}_{\ell m} $ is the angular power spectrum of the systematics. 

We plot the SFB PS $C^{S}_{\ell n_1n_2}$ of the systematics and the cosmological signals in \cref{fig:SFB_foreground_square,fig:SFB_foreground_line}, always normalized to $C^{S}_{\ell 00}$, to be agnostic to the particular type of angular systematics. With $n_1=n_2=0$, $C^{S}_{\ell 00}$ is the power spectrum at the largest radial modes for each $\ell$, so our plot illustrates the significance of the signal relative to its largest radial mode. We show the full SFB power spectrum  $C_{\ell n_1n_2}$ at a fixed $\ell=3$ in \cref{fig:SFB_foreground_square}, and we illustrate the diagonal ($n_1=n_2$) and one of the off-diagonal ($n_1=0$, $n_2\neq 0$) cross sections of different angular multipoles in \cref{fig:SFB_foreground_line}. 

In the left panels of \cref{fig:SFB_foreground_square,fig:SFB_foreground_line}, we see the cosmological signal is significant and remains at the same order of magnitude on the diagonal part of the spectra across the large and small radial scales. Generated by redshift evolution and redshift-space distortion, the off-diagonal components of the cosmological signals are significantly smaller compared to the diagonal parts but still remains at (sub)percent level of $C_{\ell 00}$ (the left panel of \cref{fig:SFB_foreground_square} and the right panel of \cref{fig:SFB_foreground_line}).

In comparison to the cosmological signals, the impact of the angular systematics is only concentrated at the largest radial mode, as evidenced by the right panel of \cref{fig:SFB_foreground_square}. Only the $n=0$ and $n=1$ modes are above the percent level of the strength of $C^{S}_{\ell 00}$, with little signal beyond $n=2$. Examining the diagonal components of the SFB PS for a fixed angular multipole in the left panel of \cref{fig:SFB_foreground_line}, we see that the angular systematics drastically drop off following $k^{-8}$ across the radial modes. The off-diagonal cross sections ($n_1=0, n_2\neq 0$) drop as $k^{-4}$, the rate of which is still significantly faster than the cosmological signal, as illustrated in the right panel of \cref{fig:SFB_foreground_line}. These observed power-law drop-offs result from the asymptotic behavior of $d_{n\ell}$, defined in \cref{eq:dnl}, which decays approximately as $k^{-4}$ for $\ell\neq 0$. Our calculations have so far assumed surveys with full-sky coverage. The presence of angular masking will mixes angular multipoles and therefore changes the angular patterns of both cosmological signals and systematics. However, the angular mixing will have minimal impacts on the radial modes, so the asymptotic drop-off behavior of $d_{n\ell}$ with respect to $k$ will hold regardless of the sky coverage.

The above analysis with the SFB basis underscores the importance of accessing radial modes in large-scale structure surveys. Even when the angular systematics dominate over the cosmological signal, cosmological information can still be recovered from smaller radial modes due to the power-law decrease of angular systematics. This is the rationale behind the design and analysis of many line intensity mapping surveys, where the angular foreground can exceed the cosmological signal by several orders of magnitude. In galaxy surveys, achieving precise redshift measurements becomes crucial, since it enables measurements of clustering at smaller radial modes, which will help to mitigate the impact of angular systematics regardless of the chosen analysis framework.

The $k^{-8}$ drop-off of the angular systematic across the radial modes demonstrates the excellent angular and radial separation offered by the SFB basis: the leakage of purely angular fluctuations to the higher radial modes is minimal. This separation offers a way to robustly control the systematics in cosmological measurements, since one can simply excise the large radial modes where the systematics cannot be accurately mitigated, or for a more refined approach, down-weight those modes appropriately. The exact number of radial modes that will need to be cut at each angular mode depends on the strength and pattern of the angular systematics, which we will illustrate using stellar contamination as a particular example in \cref{sec:star}. 

\begin{figure*}[tbp]
\centerline{\includegraphics[width=0.75\hsize]{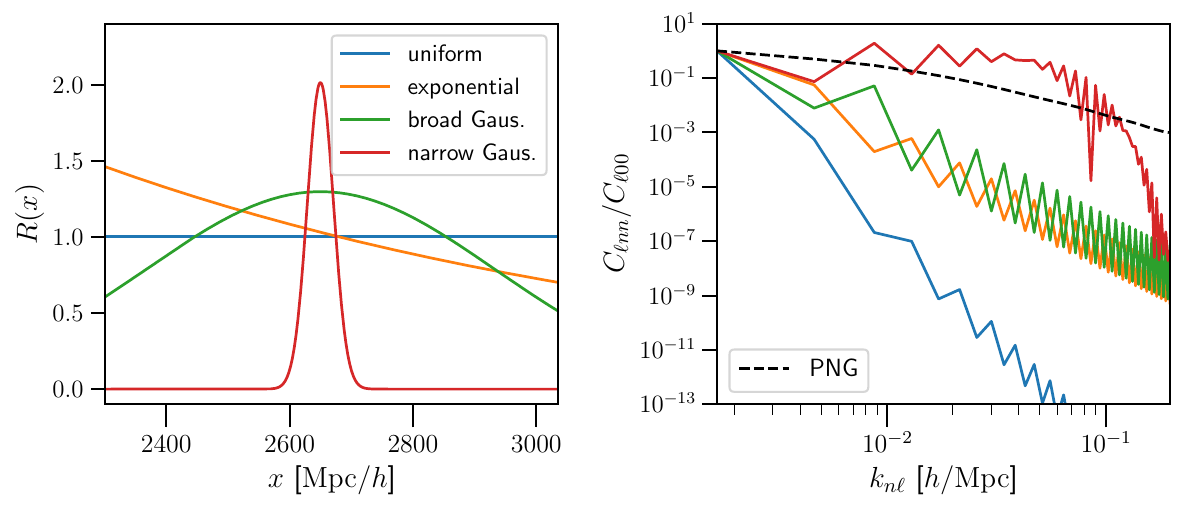}}
\caption{Comparison between the different drop-off behaviors of angular systematic under different radial distributions. The left panel shows the different radial distributions being considered, while the right panel shows the SFB PS of the systematic with these different radial distributions at $\ell=4$, normalized by the $n_1=n_2=0$ mode. In the right panel, we also plot the behavior of the local PNG signal in the solid-dashed line, and we see that for all radial distributions except the narrow Gaussian one, the systematic drops significantly faster than the PNG signal as the total Fourier mode $k$ increases.}
\label{fig:radial_distribution}
\end{figure*}

We emphasize that the angular systematic drops as $k^{-8}$, which is significantly faster than and different from the signals imprinted by local primordial non-Gaussianities (PNG) or general relativistic effects, which generally drop as $k^{-2}$ across the angular modes of the SFB PS \cite{24GR_SFB}, as shown in the right panel of \cref{fig:radial_distribution}. Therefore, an angular systematic effect will not contaminate the PNG signal in the radial modes. By doing a pattern search that differentiates the $k^{-2}$ signal of interests from the $k^{-8}$ signal of systematics, the angular systematics may only leave the largest radial modes with $n=0$ unusable even if the angular systematic can potentially dominate over the clustering signal for many more radial modes.

We have so far assumed a uniform selection function within the redshift bin, that is assuming a constant angular systematic across the redshift. In reality, there could be a radial dependence $R(x)$ coming from a radial selection function or the radial evolution of the angular systematic (while still considering an additive systematics that is separable in angular and radial components). In this case, we have
\begin{align}
    S_{n \ell m}
&=S_{\ell m}\int_{x_{\rm min}}^{x_{\rm max}}\dd x\,x^2g_{n\ell}(x)R(x)\,,
\end{align}
where the integrand in \cref{eq:dnl} for the $d_{n\ell}$ constant is now modulated by the radially dependent function $R(x)$. 
To study the possible impact of different modulations, we plot in  \cref{fig:radial_distribution} the effects of a few representative radial distributions $R(x)$ on the SFB PS of the angular systematic. All the radial distributions considered are given in terms of comoving radial distances. We see that a uniform radial distribution has the steepest drop-off as $k^{-8}$, while the SFB PS for an exponential or a broad Gaussian distribution falls as $k^{-4}$, still leaving the smaller radial modes mostly uncontaminated. The systematic under these broad radial distributions will not be confused with a PNG signal due to their different radial patterns. 

In contrast, the PS of a narrow Gaussian distribution (the red line of \cref{fig:radial_distribution}) falls at a shallower rate and has more resemblance to a PNG signal in SFB space. A more radially localized systematic can contaminate significantly more radial modes, and the strategy of cutting $n=0$ modes becomes less effective in such cases. Fortunately, it is possible to detect such localized systematics in real space by comparing cosmological analyses at different redshift ranges within a redshift bin. Even though a more radially localized distribution (more similar to the Dirac-delta function) becomes more spread out in SFB space, these systematics can still be directly mitigated in real space due to their localized features there. 

As we demonstrate in \cref{fig:radial_distribution}, the radial distribution of the systematic impacts its particular pattern in SFB space, so it is generally important to develop systematics templates in 3D space. Almost all previous LSS analyses rely on 2D systematics templates in the angular space, which can be insufficient to distinguish cosmological signals from systematics under the precision requirements of the upcoming surveys.

\subsection{Radial systematics}\label{sec:radial-foreground}
For completeness, we next provide a brief study on the SFB signal of radial systematics. Most radial systematics in galaxy surveys, including galaxy radial selection and redshift measurement errors, are multiplicative or convolutional\footnote{Our systematics model of \cref{eq:sys} do not account for systematics, such as the redshift measurement uncertainties, that are convolutional in configuration space.} in nature. For the additive component of a radial systematic effect $S^{\rm R}(x)$, the SFB mode is
\begin{align}
    S^{\rm R}_{n\ell m}
&=\int_{x_{\rm min}}^{x_{\rm max}}\dd x\,x^2g_{n\ell}(x)S^{\rm R}(x)\int \dd^2{\hat{\bn}}\,Y^*_{\ell m}(\hat{\bn})\nonumber\\
&=S_{n\ell}^{\rm R}\delta^{\rm K}_{\ell0}\delta^{\rm K}_{m0}\sqrt{4\pi}\,,
\end{align}
where we have used \cref{eq:Y_mean}. Therefore, the SFB PS of the radial systematic is
\begin{align}
    C_{\ell n_1 n_2}^{S,{\rm R}}=\frac{4\pi}{2\ell+1}\delta^{\rm K}_{\ell 0}S_{n_1\ell}^{\rm R}S_{n_2\ell}^{\rm R}\,.
    \label{eq:radial-sys-sfbps}
\end{align}
The above equation shows that the additive component of a radial systematic can be entirely isolated to the modes at the angular monopole in the full-sky limit. Therefore, removing the angular monopole will ensure the cosmological measurements to be robust against unknown, purely radial systematics.

This is indeed the motivation behind the radial integral constraint (RIC) \cite{19Mattia_IC,24_DESI_Y1_PNG}, already used in galaxy surveys to mitigate uncertainties in knowledge of radial selection function. Rather than modeling the radial selection function directly, the observed galaxy redshift distribution—including both cosmological anisotropies and selection effects—was directly enforced into the random catalogs used for power spectrum estimation. This procedure, also known as the local average effect \cite{12dePutter_LAC,20Wadekar_covariance,23Gebhard_SFB_eBOSS}, circumvents the need for explicit modeling of the radial selection function. In the full-sky case, the RIC exactly removes all components of the $\ell=0$ mode of the SFB power spectrum, where purely radial systematics reside as shown in \cref{eq:radial-sys-sfbps}. For treatments of RIC in the SFB basis under realistic sky masks, we refer readers to Ref.~\cite{23Gebhard_SFB_eBOSS,24Wen_gSFB}.

\subsection{Stellar contamination}\label{sec:star}
We next give an example of the SFB power spectrum computed using a slightly more realistic 3D template. We choose stellar contamination since it is predominantly an additive effect on the galaxy number density, so our consideration of only additive effect on the density contrast is accurate. Stellar contamination can contribute significantly to the excess of correlation at the large angular scales for tracers such as quasars \cite{22Chaussidon_DESILS_QSO}.  

For the angular distribution of stars, we use the density of point sources across the full sky from 
\textit{Gaia} Data Release 3 \cite{23Gaia_DR3} in the magnitude range: $12 < \texttt{PHOT\_G\_MEAN\_MAG} < 17$. This is similar to the stellar templates used in current galaxy surveys \cite{22Chaussidon_DESILS_QSO,24_DESI_Y1_PNG}, and the magnitude cut represents the range where the \textit{Gaia} survey is essentially complete \cite{18Gaia_DR2}. The angular power spectrum of the above stellar template can be roughly approximated by a power law 
\begin{equation}C_{\ell}^{\mathrm{stellar}}\sim (\ell+1)^{-2.3},
\end{equation}
which is also similar to the scaling of the angular power spectrum of the Galactic dust \cite{16Planck_polarized_dust}. Therefore, stellar contamination is representative of systematics of Galactic origins. For the radial distribution, we simply assume the redshift of the stars (that are misclassified as galaxies) is uniformly distributed across the redshift bin.

\begin{figure}[tbp]
\centerline{\includegraphics[width=1.05\hsize]{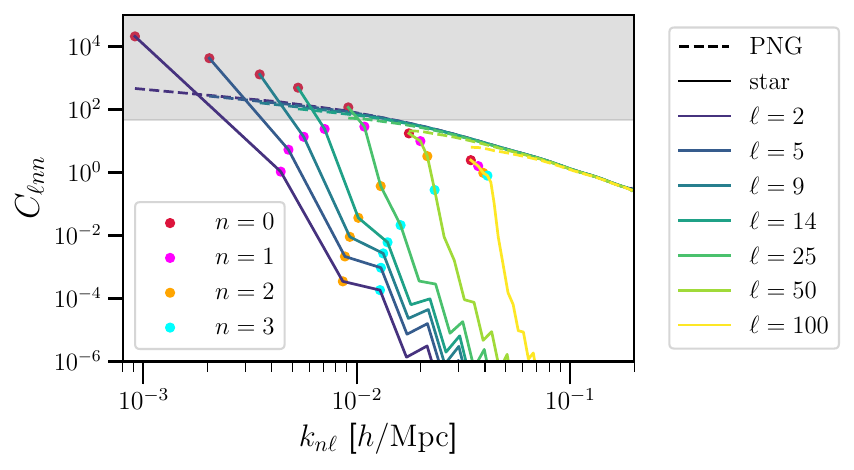}}
\caption{Comparison between stellar contamination (with contamination fraction 2.5$\%$) and PNG signals (with $\fNL=1$, and we show the difference of the SFB PS calculated with $\fNL=1$ and $\fNL=0$) for the diagonal components of the SFB power spectrum for $z=1.0-1.5$. The solid lines represent the stellar foreground, while the dashed lines represent the PNG signal at different angular multipoles $\ell$. We use the colored points to show the radial indices $n$ for the stellar foreground. The stellar contamination only exceeds the PNG signals at the largest radial modes with $n=0$ and the large angular scales with $\ell\lessapprox 50$ (roughly corresponding to the modes in the horizontal grey band). The stellar foreground is subdominant compared to the PNG signal at other modes. }
\label{fig:SFB_stellar}
\end{figure}

\begin{figure}[tbp]
\centerline{\includegraphics[width=0.85\hsize]{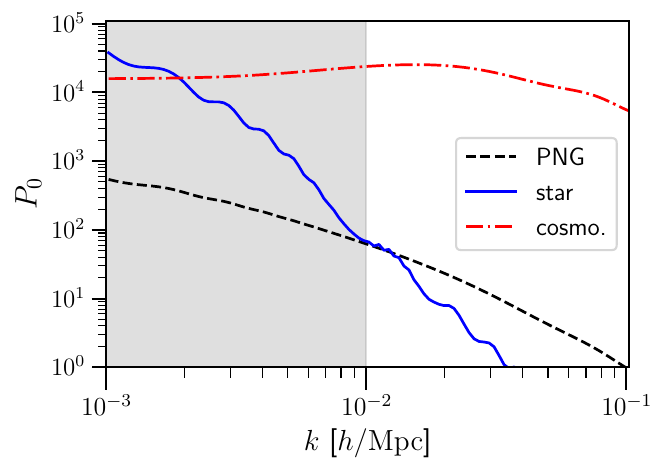}}
\caption{Comparison between the PNG (dashed line), stellar contamination (solid line), and cosmology (dashed-dot line) signals for the power spectrum monopole, using the same settings as \cref{fig:SFB_stellar}. Both PNG and stellar foregrounds have scale-dependent behaviors, and stellar foregrounds dominate over PNG signals at $k\lesssim0.01 h/{\rm Mpc}$ (indicated by the region of the grey band), suggesting the loss of all the large-scale modes due to stellar contamination.}
\label{fig:P0_stellar}
\end{figure}

Assuming a stellar contamination fraction of $2.5\%$ to the observed galaxies, we show the SFB PS  of both stars and the local PNG signal of $f_{\rm NL}=1$ (the difference of the SFB PS between $\fNL=1$ and $\fNL=0$ with all other parameters fixed) in \cref{fig:SFB_stellar}. We use colored lines to indicate angular multipoles $\ell$, and the different colored dots to illustrate the first few radial modes $n$. We see that the amplitude of the PNG signal only slightly decreases for smaller angular and radial scales, while the PS of stars decreases drastically for both smaller angular and radial scales. In particular, the stellar contamination predominantly impacts the lower $\ell$ portions $(\ell<50)$ of the $n=0$ modes, suggesting the effect of stellar contamination is localized in SFB space. As a result, only part of the $n=0$ modes need to be removed due to stellar systematics dominating over the PNG signal, while other large radial modes with $n=1$ and $2$ can still be used for PNG measurements. Varying the stellar contamination fraction and the $\fNL$ parameter will alter the amplitudes of the stellar and PNG power spectra, thereby changing which radial modes can be used for measuring PNG.

With a known angular distribution, we can now also compute the effects of stellar foregrounds in the Cartesian power spectrum monopole, the clustering statistics used in most PNG analyses to date. We can compute the PS monopole from the SFB PS using the SFB-to-PSM mapping developed in Refs.~\cite{24Wen_gSFB,24PSM_SFB}, and the results for the PNG ($f_{\rm NL}=1$), stellar, and cosmologiucal signals are shown in \cref{fig:P0_stellar}. We use the smooth power-law $(\ell+1)^{-2.3}$ instead of the measured, noisy angular power spectrum of \textit{Gaia} stars in the computation such that the PS monopole of stars will remain relatively smooth and well-behaved. 

The PNG and star have a similar scale-dependent pattern in the PS monopole, which is in stark contrast to the two signals having completely different angular and radial dependence in SFB space as shown in \cref{fig:SFB_stellar}. Stellar foregrounds dominate over PNG signals at $k\lesssim0.01 h/{\rm Mpc}$, which suggests the need to remove all the large-scale modes with $k\lesssim0.01 h/{\rm Mpc}$ in the PS monopole. In comparison, in an SFB-based analysis, one only needs to remove the $n=0$ modes, while the $n=1$ and $n=2$ modes with $k\lesssim0.01 h/{\rm Mpc}$ can still be retained due to the PNG signal being stronger than the stellar foreground at these modes. In an analysis based on PS monopole, these $n=1$ and $n=2$ modes will be discarded due to the unavoidable mixing of angular and radial information in the Fourier space, enforcing a strict cut of modes below $k_{\rm min}$. Since most constraining power of local PNG comes from the large scales, the unnecessary loss of the large radial modes can lead to a significant degradation in the $f_{\rm NL}$ constraint. This example illustrates the power of SFB PS in localizing systematics and retaining more uncontaminated modes compared to a PSM-based analysis.

\section{Discussion}\label{sec:discussion}

\subsection{Systematics mitigation via SFB}\label{sec:mitigation}

With the advent of the SFB basis as a feasible choice for analyzing current LSS data, we gain access to a powerful framework for mitigating observational systematics in a more flexible and interpretable manner. Building on the above examples, we now outline a broader strategy for systematics mitigation enabled by this basis. 

When the templates of known systematics are accurately characterized, their impacts on the data can be effectively mitigated using established techniques such as template subtraction \cite{11Ross_template_SDSS,12Ho_SDSS,16Elsner_template}, mode (de)projection \cite{92Rybicki_deprojection,13Leistedt_mode}, or regression-based methods \cite{20Weaverdyck_systematics_review,17Ross_BOSS_systematics,18Poole_DESY1,23_DESI_LRG,20Rezaie_DECaLS_systematic}. In such idealized situations where systematics are completely known and can be cleaned, all modes can be used for cosmological inference. 

In practice, however, there are always uncertainties in our knowledge of systematics, and the impacts of unknown systematics pose serious challenges for any cosmological measurements. This is where the SFB formalism offers a unique advantage in avoiding residual or unknown systematics, since many of them are localized in specific angular and radial regions of SFB space. As we previously discussed, systematics with broad radial distributions predominantly impact the largest radial modes, so one can cut the $n=0$ modes from the analysis to be robust against any systematics with broad radial distributions. 

In general, one can choose particular angular and radial modes to be excluded from the analysis based on where the particular systematics of concern are localized in SFB space. The SFB basis enables surgical removal of specific angular and radial modes affected by systematics, without discarding all the statistical power at a given scale—as opposed to the $k_{\rm min}$ cuts in PSM analyses (as seen in \cref{sec:star}). Moreover, one can test the consistency of inference results based on different angular and radial cuts in the SFB framework, which will allow one to assess the robustness of the cosmological measurements against unknown systematics.

Eliminating modes dominated by uncertain systematics is akin to assigning each mode a weight based on how well we understand its underlying systematics. Modes with perfectly understood systematics can be fully retained, while those with completely unknown systematics are best discarded. However, there can also be a middle ground: modes with partial or imperfect knowledge might be down-weighted rather than entirely removed, thereby preserving some of their informational value. Ultimately, these weights in SFB space will reflect the extent to which we believe the cosmological signal is contaminated by systematics, which requires a certain level of confidence in estimating the level of systematics.

Besides down-weighting particular SFB modes according to systematic priors at the estimator level, one can also jointly infer the systematics and cosmological parameters, provided there is enough prior knowledge to parametrize the systematics. In particular, if the systematic component factorizes into angular and radial parts, one can parametrize the SFB PS and marginalize over these parameters to obtain cosmological constraints. This strategy mirrors aspects of CMB analyses, where the angular power spectra of foregrounds are parametrized and the foreground parameters jointly inferred with the cosmological parameters \cite{18Planck_likelihood,21BK18}. Modeling and marginalizing over systematics in Cartesian Fourier space will be more cumbersome due to the mismatch with the geometry of observations and foregrounds, whereas the SFB basis naturally accommodates angular and radial templates for systematics, making the construction of a parametric model for systematics considerably more straightforward. 

These capabilities to localize, excise, down-weight, and parametrize systematics at a mode level showcase the substantial potential of the SFB framework in 3D clustering analyses. By matching the geometric structures of both observational and astrophysical systematics, the SFB basis allows one to draw on the success of mitigation methods used in CMB analyses and opens up new avenues for systematic treatments in LSS surveys.

\subsection{Comparison with systematics mitigation via clustering wedge}\label{sec:compare-wedge}

Establishing the traditionally used $P(k,\mu)$ as the plane-parallel limit of the SFB PS has important implications: we can now understand the success of clustering wedge approaches at systematics mitigation as coming from radial and angular mode separation, which the SFB basis can now extend properly to the fully curved sky, transferring tools and strategies from the clustering wedge, while enabling the use of a single estimator across all scales. Here we offer a detailed comparison between the SFB PS and the clustering wedge in the context of systematics mitigation. 

The clustering wedge $P(k,\mu)$ (or the cylindrical power spectrum $P(\kpe,\kpa)$) has long been recognized as a powerful space for diagnosing and mitigating systematics in the linear and quasi-linear regime of LSS surveys. Examples of such systematics include fiber collisions in spectroscopic galaxy surveys \cite{17BOSS_fourier_wedge,17Pinol_wedge,17Hand}, astrophysical foregrounds \cite{20Cunnington_IM_wedge_foreground,20Cunnington_foreground_HI} and interferometric fringes \cite{14Liu_wedge1,14Liu_wedge2,25Munchi_interferometric} in line intensity mapping surveys, and interloper contamination in both types of LSS surveys \cite{16Lidz_interloper,16Cheng_interloper,19Gebhardt_interloper}. These systematics often produce particular patterns or can be localized in certain regions of the $(\kpe,\kpa)$ space, usually at the low $k_{||}$ (i.e., $\mu\approx 0$), which correspond to predominantly angular modes.

As shown in \cref{sec:numerical-approx}, the clustering wedge emerges as the flat-sky limit of the SFB PS: the angular and radial modes in the SFB PS map to the transverse and LOS Fourier modes in the clustering wedge at the small-scale limit. Consequently, systematics that localize cleanly in SFB space also remain localized in the clustering wedge. However, the transverse and LOS modes are no longer well-defined for the largest scales with large angular separations, while the SFB basis retains its interpretability and remains effective for systematics localization at the wide-angle regime, as we have demonstrated in \cref{sec:additive-syst}. This makes the SFB basis the natural and preferred choice for systematic mitigation at the largest scales.

Even for quasi-linear scales where the two formalisms are analogous, there are still reasons to favor the SFB formalism over the clustering wedge approach regarding the implementation of estimators and the binning of data vectors. Although the clustering wedge is straightforward to estimate in small-field surveys with a well-defined global LOS, wide-field surveys pose a greater challenge. With the Yamamoto estimator, it is possible to construct the clustering wedge based on the observed PSM for a wide-field survey through \cite{17Hand,17BOSS_fourier_wedge}
\begin{align}
    \hat{P}(k,\mu_i)=\sum_{L}\hat{P}_L(k)\bar{\mathcal{L}}_{L}(\mu_i)\,,
    \label{eq:wedge}
\end{align}
where $\bar{\mathcal{L}}_{L}(\mu_i)\equiv \int_{\mu\in\mu_i}\dd\mu\,\mathcal{L}_{L}(\mu)$ is the average of the Legendre polynomials in the corresponding $\mu$ bins. Under linear theory, $L=0,2,4$ are all the multipoles needed for measuring the cosmological components of the clustering wedge from data. 

However, the necessary $L_{\rm max}$ can easily reach $O(10)$\footnote{For example, Ref.~\cite{17Hand} advocates for $L_{\rm max}=16$ for mitigating the effects of fiber collision.} to account for certain systematics, significantly higher than the $L_{\rm max}=4$ used for cosmological analyses. In such cases, the FFT-based Yamamoto estimator loses much of its computational advantage over the SFB PS estimator, since working at large $L_{\rm max}$ can become significantly more expensive. Here, the SFB-based estimator provides a practical alternative for estimating the clustering wedge. The SFB PS estimator such as \texttt{SuperFaB} \cite{21Gebhardt_SuperFab} can be deployed up to the quasi-linear scales at $0.2\,h/{\rm Mpc}$ where the theoretical modeling of the power spectrum remains robust. Even though both estimators are deployable at these quasi-linear scales, the SFB estimator is more agnostic to the structure of systematics, making it the more flexible choice, whereas one has to determine the $L_{\rm max}$ needed for each type of systematics in order for the constructed wedge to capture their angular patterns. 

Another practical benefit of the SFB estimator lies in its natural discretization. Constructing $P(k,\mu)$ requires choosing arbitrary binning schemes in $k$ and $\mu$ due to their inherently continuous nature, which can affect how cleanly systematics localize and may lead to additional leakage if poorly chosen. In contrast, the SFB decomposition provides a complete, non-redundant set of orthonormal modes for a finite-volume survey, and captures systematic effects at particular angular and radial modes. This eliminates ambiguity in the mode localization of systematics. Since the relationship between $P(k,\mu)$ and SFB PS is well-established through \cref{eq:pp-SFB} at quasi-linear scales, the SFB estimator effectively provides an estimate for a discrete version of the clustering wedge, where $k$ and $\mu$ are properly discretized according to the survey volume. 

In summary, the SFB basis offers advantages for systematics mitigation at the largest scales due to its better match with the spherical nature of large-area surveys, and it can still be the preferred choice at the quasi-linear scales compared to clustering wedge for practical considerations. Consequently, an SFB-based analysis allows one to control and mitigate systematics across all scales with a single estimator.

\section{Conclusion}\label{sec:conclusion}

The unprecedented statistical power of the upcoming LSS surveys demands exquisite control over observational systematics, in order to achieve unbiased cosmological measurements. However, the Cartesian power spectrum multipoles, which have been used in most of the large-scale analyses to date, become increasingly inconsistent with the spherical geometry of the light cone, as surveys cover a wider angular area and broader redshift range. This mismatch between the basis and survey geometry causes challenges for both theory and systematics where angular and radial modes are non-trivially mixed in the Cartesian Fourier basis. These challenges strongly motivate the use of the spherical Fourier-Bessel basis, where the angular and radial modes are properly discretized and separated into the two indices $\ell$ and $n$, as we discussed in \cref{sec:radial}.  

Through Eq.~\ref{eq:pp-SFB}, we approximated the SFB power spectrum with the Cartesian power spectrum under the plane-parallel limit. This connection with the Cartesian framework helps us better understand the large data vectors in the SFB PS. Since the SFB PS is the full-sky generalization of the clustering wedge, we expect most of the previous studies on systematics based on the Cartesian clustering wedge $P(k,\mu)$ to extend to the SFB PS. While the $(k_\parallel, k_\perp)$ modes used in the wedge formalism provide a useful separation of radial and angular scales, it is only valid under the flat-sky approximation for surveys with small angular coverage and cannot be consistently applied to wide-field surveys. In contrast, the SFB power spectrum maintains exact separation of angular and radial modes under curved-sky geometry, and can be directly and efficiently measured on the full sky, making it an ideal framework for systematic mitigation in LSS analyses.

Using a generic additive angular systematic effect, with stellar contamination as a specific example, we have demonstrated the SFB basis' distinct advantage in isolating systematics. Systematics with broad and smooth radial distributions predominantly localize in the $n=0$ modes, allowing other large-scale radial modes to be preserved for cosmological analysis. In contrast, standard Cartesian 3D clustering analyses of wide-field survey based on power spectrum multipoles must discard all modes below a given 
$k_{\rm min}$ because they cannot distinguish between angular and radial contributions. The ability of the SFB framework to cleanly separate these contributions on the full sky—not just in the flat-sky limit—is what enables more targeted mitigation strategies and distinguishes it from Cartesian approaches.

The SFB basis is particularly promising for measuring primordial non-Gaussianity and relativistic  effects, as radially smooth systematics exhibit distinct patterns in SFB space compared to the characteristic 
$k^{-2}$ scaling of PNG and GR signals. These findings highlight the SFB power spectrum as a powerful tool of constraining fundamental physics at the largest scales while mitigating over uncertain or unknown systematics. 

Furthermore, an SFB-based analysis is also complementary and beneficial to the standard PSM-based analysis, since PSM can be viewed as a weighted-sum compression of the SFB PS \cite{24PSM_SFB,24Wen_gSFB}. One can therefore also use the SFB power spectrum as an intermediate step to handle systematics and subsequently use the compressed quantities—such as the PSM—for further inference.

Our current studies have so far relied on theoretical calculations under simplified assumptions about systematics, that is the systematics is either predominantly angular and radial. It will be interesting to further develop the technique and examine more realistic and survey-specific 3D templates of systematics under the SFB basis. The goal will be to demonstrate the procedure of controlling and mitigating systematics with the discrete SFB basis on cosmological data in subsequent work. 

Our results can also be extended to the higher-order statistics. The clear separation of angular and radial scales in the SFB basis offers a promising new avenue for systematic mitigation in three-point clustering statistics. Our results strongly motivate further developments in the SFB bispectrum \cite{18Bertacca_GR_SFB_Bi,23Benabou_SFB} for constraining fundamental physics and achieving robust cosmological measurements beyond the two-point statistics.

\section*{Acknowledgements} 
We thank Sean Bruton for providing the stellar templates based on \textit{Gaia}. We thank Jamie Bock, Yongjung Kim, Richard Feder, James Cheshire, Yun-Ting Cheng and the rest of the SPHEREx cosmology team for useful discussions, comments and feedback. We acknowledge support from the SPHEREx project under a contract from the NASA/GODDARD Space Flight Center to the California Institute of Technology. Part of this work was done at Jet Propulsion Laboratory, California Institute of Technology, under a contract with the National Aeronautics and Space Administration (80NM0018D0004). RYW further acknowledges support through the Canada Graduate Research Scholarship – Doctoral program (CGRS D) from the Natural Sciences and Engineering Research Council of Canada (NSERC).

\appendix

\section{Analytical Plane-parallelization}\label{sec:analytical-approx}

We here discuss how the angular and radial modes in the SFB basis can be analytically approximated in the flat-sky limit. The SFB mode defined in \cref{eq:sfb_discrete_fourier_pair_b} is the radial transform of the spherical harmonic mode over the comoving distance:
\begin{align}
    \delta_{n\ell m}=\int_{x_{\rm min}}^{x_{\rm max}}\dd x\,x^2g_{n\ell}(x)\delta_{\ell m}(x)\,,
    \label{eq:SFB-TSH}
\end{align}
where the spherical harmonic transformation of an angular field and its inverse are given as
\begin{align}
\delta_{\ell m}&=\int \dd^2\hat{\bx}\,Y^*_{\ell m}(\hat{\bx})\delta(\hat{\bx})\,,\label{eq:TSH-transform}\\
\delta(\hat{\bx})&=\sum_{\ell=0}^{\infty}\sum_{m=-\ell}^{\ell} Y_{\ell m}(\hat{\bx})\,\delta_{\ell m}\,.\label{eq:TSH-transform-inverse}
\end{align}

Following Refs.~\cite{97Zaldarriaga_CMB,99White_CMB,00Hu_CMBLens,23Gao_flat_angular}, we first review how the spherical harmonic decomposition can be approximated by a 2D Fourier transform in the plane perpendicular to the LOS. We then examine the radial basis functions in the SFB basis and show how they can be approximated by plane waves in the radial direction. Putting the angular and radial approximations together yields an approximation to the SFB power spectrum, of which the numerical implementation has been given in \cref{sec:numerical-approx}.

\subsection{Angular modes}\label{sec:analytical-angular}

We first define the following weighted sum of the spherical harmonic moments
\begin{align}
    \delta({\bold l})\equiv\sqrt{\frac{4\pi}{2l+1}}\sum_{m}i^{-m}\delta_{\ell m}e^{im\phi_{\ell}}\,\label{eq:pp-l},
\end{align}
where ${\bold l}$ is the vector of length $\ell$ and angle $\phi_{\ell}$ in the polar coordinates. The inverse of the above transform is
\begin{align}
    \delta_{\ell m}=\sqrt{\frac{2l+1}{4\pi}}i^{m}\int \frac{\dd\phi_l}{2\pi}e^{-im\phi_{\ell}}\delta({\bold l})\,.\label{eq:pp-l-inverse}
\end{align}
We will now show $\delta({\bold l})$ as defined above is the 2D Fourier modes in the flat-sky case. For small angles around the pole, spherical harmonics can be approximated as
\begin{align}
    Y_{\ell m}(\hat{\bx})\approx \sqrt{\frac{\ell}{2\pi}}J_{m}(\ell\theta)e^{im\phi}\,,\label{eq:Ylm-approx}
\end{align}
where $J_m$ is the $m$-th Bessel function of the first kind, and the full-sky 3D spherical coordinates $\hat{\bx}=(\theta,\phi)$ are approximated by the flat-sky 2D polar coordinates $\boldsymbol{\theta}\equiv(r \approx \theta,\phi)$ on the plane perpendicular to the LOS.

Substituting \cref{eq:pp-l-inverse,eq:Ylm-approx} into Eq.~\ref{eq:TSH-transform-inverse}, we have
\begin{align}
\delta(\hat{\bx})&\approx \sum_{\ell} \sqrt{\frac{\ell(2\ell+1)}{8\pi^2}}\int \frac{\dd\phi_l}{2\pi}\delta({\bold l})\sum_m i^{m}e^{im(\phi-\phi_{\ell})}J_{m}(\ell\theta)\nonumber\\
&=\sum_{\ell} \sqrt{\frac{\ell(2\ell+1)}{8\pi^2}}\int \frac{\dd\phi_l}{2\pi}e^{i\boldsymbol{l}\cdot\boldsymbol{
    \theta
    }}\delta({\bold l})\nonumber\\
&\approx \int \frac{\dd^2{\bold l}} {(2\pi)^2}e^{i\boldsymbol{l}\cdot\boldsymbol{
    \theta
    }}\delta({\bold l})\,\label{eq:2D-Fourier-inverse},
\end{align}
where we have applied the 2D plane-wave expansion of \cref{eq:JA-expansion} and approximated the discrete sum over $\ell$ with an integral. The above equation establishes $\delta({\bold l})$ as the 2D Fourier-transform of the 2D angular field
\begin{align}
\delta({\bold l})\approx  \int \dd^2\hat{\bx}\,e^{-i\boldsymbol{l}\cdot\boldsymbol{
    \theta}}\delta(\hat{\bx})\,,
    \label{eq:2D-Fourier-transform}
\end{align}
that is $\boldsymbol{l}$ as the Fourier conjugate to the flat-sky 2D coordinates $\boldsymbol{\theta}$ and therefore the flat-sky angular modes. \Cref{eq:pp-l,eq:pp-l-inverse} show how the flat-sky and full-sky angular modes are related. The flat-sky approximation generally holds for smaller angular scales corresponding to larger $\ell$ values.

The flat-sky 2D polar coordinates $\boldsymbol{\theta}$ represent the angles in the full-sky 3D spherical coordinates and are therefore dimensionless. To relate the angle coordinates with the comoving distance coordinates $\bx_{\perp}$ on the plane perpendicular to the LOS at distance $x$ away from the observer, we have $\bx_{\perp}\equiv x\boldsymbol{\theta}\approx(x\theta,\phi)$ in the polar coordinates. The transverse Fourier mode on the perpendicular plane is then related to the flat-sky angular mode $\boldsymbol{l}$ as
\begin{align}
    \bk_{\perp}=\frac{\boldsymbol{l}}{x}\,
    \label{eq:angular_approx},
\end{align}
and the transverse Fourier transform is
\begin{align}
\delta(\bk_{\perp})=\int \dd^2{\bx_{\perp}}\, e^{-i\bk_{\perp}\cdot\bx_{\perp}}\delta(\bx_{\perp})=x^2\delta(\boldsymbol{l})\,.
\label{eq:perpendicular-Fourier-transform}
\end{align}

\subsection{Radial modes}\label{sec:analytical-radial}
We now examine the analytical approximation to the radial basis functions in the SFB basis. Since the SFB basis consists of the eigenfunctions of the Laplacian, the radial basis functions satisfy the radial part of the Helmholtz equation (the Laplacian eigenequation)
\begin{align}
\frac{\dd}{\dd x}\left(x^2\,\frac{\dd g_{\ell}(kx)}{\dd x}\right)
+ \big[(kx)^2 - \ell(\ell+1)\big]\,g_{\ell}(kx)=0\,.
\label{eq:helmholtz-radial}
\end{align}

In the limit of $(kx)^2\gg \ell(\ell+1)$, the effects of the angular eigenvalues become negligible, and the Fourier $k$ modes are dominated by the radial modes. The radial basis function $g_{\ell}(kx)$ can then be approximated with functions $g_{\rm p}(kx)$ that solve
\begin{align}
    \frac{\dd}{\dd r}\left(x^2\,\frac{\dd g_{\rm p}(kx)}{\dd x}\right)
+ (kx)^2 g_{\rm p}(kx)=0\,.
\label{eq:helmholtz-radial-pp}
\end{align}
One can verify that 
\begin{align}
    g_{\rm p}(kx)= \frac{1}{x}A\cos(k(x-x_0))
    \label{eq:gp-kx}
\end{align}
is the general solution to the differential equation of \cref{eq:helmholtz-radial-pp}. Note that when $\ell=0$, \cref{eq:gp-kx} gives the exact solution to the radial Helmholtz equation
\begin{align}
    g_{0}(kx)=g_{p}(kx)\,.
    \label{eq:g0}
\end{align}

Assuming $x_{\rm max}$ and $x_{\rm min}$ are reasonably close (that is $\Delta x\equiv x_{\rm max}-x_{\rm min}\ll x_{\rm min}$), we can ignore the variation of $1/x$ term and treat it as a constant at some effective distance $x_{\rm eff}$ in the redshift bin.  The velocity boundary conditions of \cref{eq:velocity-boundary-def} and the normalization requirement of \cref{eq:gnl_orthonormality} lead to the following discretized radial basis functions
\begin{align}
  g_{\ell}(k_nx)&\approx g_{\rm p}(k_nx)\nonumber\\
  &\approx\frac{1}{x_{\rm eff}}\sqrt{\frac{2}{\Delta x}}\cos(k_n(x-x_{\rm min}))\label{eq:gnl-cosine-approx}\,,
\end{align}
and the quantization of $k$ modes yield
\begin{align}
      k_n = \frac{n\pi}{\Delta x}\approx k_{||}\,,
     \label{eq:radial_approx}
\end{align}
where we have assumed the radial modes to be the dominant contribution in the total Fourier modes. 

Even though the cosine form in \cref{eq:gnl-cosine-approx} is obtained assuming $(kx)^2\gg \ell(\ell+1)$, it in fact provides good approximation to the radial basis function under any $n,\ell$ (in the region between the transition distance $x^{\rm t}_{n\ell}$ and $x_{\rm max}$), if one adopts $\Delta x\equiv x_{\rm max}-x^{\rm t}_{n\ell}$ and the effective distance defined in \cref{eq:eff-xnl} for each mode. Therefore, \cref{eq:gnl-cosine-approx} effectively applies to any SFB mode.

We can now approximate the SFB radial basis functions in terms of the 1D plane wave
\begin{align}
g_{\ell}(k_nx)\approx \frac{1}{x_{\rm eff}}\sqrt{\frac{2}{\Delta x}}{\rm Re}[e^{ik_n(x-x_{\rm min})}]\,,
\end{align}
and the radial SFB transform can be approximated with a Fourier transform in the radial directions
\begin{align}
    \delta_{n\ell} &= \int_{x_{\rm min}}^{x_{\rm max}}\dd x\,x^2g_{n\ell}(x)\delta(x)\nonumber\\
    &\approx\frac{1}{x_{\rm eff}}\sqrt{\frac{2}{\Delta x}}{\rm Re}[\int_{x_{\rm min}}^{x_{\rm max}}\dd x\,x^2e^{ik_n(x-x_{\rm min})}\delta(x)]\nonumber\\
    &\approx\frac{1}{x_{\rm eff}\sqrt{2\Delta x}}{\rm Re}[\int_{-\infty}^{\infty}\dd x\,x^2e^{ik_{||}x}\delta(x)]\,,
    \label{eq:radial-transform-approx}
\end{align}
where we have extended the integration range to the entire 1D real line by assuming $\delta(\bx)$ to vanish outside the SFB volume and allowing $x$ and  $k_{||}$ to take negative values. 

\subsection{Power spectrum}\label{sec:analytical-PS}

Combining the angular approximations of \cref{eq:pp-l-inverse} and \cref{eq:perpendicular-Fourier-transform} along with the radial approximation of \cref{eq:radial-transform-approx}, the SFB mode in \cref{eq:SFB-TSH} becomes
\begin{align}
    &\delta_{n\ell m}\approx\frac{1}{x_{\rm eff}}\sqrt{\frac{2l+1}{8\pi\Delta x}}i^{m}\int\frac{\dd\phi_l}{2\pi}\,e^{-im\phi_{\ell}}\nonumber\\
    &\qquad\qquad {\rm Re}\left[\int_{-\infty}^{\infty}\dd x\ e^{ik_{||}x} \delta(\bk_{\perp},x)\right]\nonumber\\
    &\approx \frac{1}{x_{\rm eff}}\sqrt{\frac{l}{4\pi\Delta x}}i^{m}\int\frac{\dd\phi_l}{2\pi}e^{-im\phi_{\ell}}{\rm Re}[\delta(\bk,x_{\rm eff})]\label{eq:SFB-mode-approx}\,,
\end{align}
where we consider the Fourier-space overdensity field to be measured at some effective distance, and combining \cref{eq:angular_approx,eq:radial_approx} give the 3D Fourier mode
\begin{align}
  \bk\equiv\left(\bk_{\perp}=\frac{\boldsymbol{\ell}}{x_{\rm eff}},k_{||}= \pm\frac{n\pi}{\Delta x}\right)
    \label{eq:3D-k-n}
\end{align}
that depends upon $\phi_{\ell}$.

We now approximate the SFB power spectrum:
\begin{align}
    &C_{\ell n n'}=\langle\delta_{n\ell m}\delta^*_{n'\ell' m'}\rangle\nonumber\\
    &\approx\frac{1}{x_{\rm eff}^2}\frac{\sqrt{\ell\ell'}}{4\pi \Delta x}i^{m-m'}\int\frac{\dd\phi_l}{2\pi}\int\frac{\dd\phi'_l}{2\pi}e^{-im\phi_{\ell}}e^{im'\phi_{\ell}'}\nonumber\\
    &\qquad{\rm Re}[\langle\delta(\bk,x_{\rm eff})\delta^*(\bk',x_{\rm eff})\rangle]\nonumber\\
    &\approx\frac{1}{x_{\rm eff}^2}\frac{\sqrt{\ell\ell'}}{2\Delta x}i^{m-m'}\int \dd\phi_l\,\int \dd\phi'_l\;e^{-im\phi_{\ell}}e^{im'\phi_{\ell}'}\nonumber\\
    &\qquad P(k,\mu,x_{\rm eff})\delta^{\rm D}(\bk-\bk')\,,
\end{align}
where we have assumed the translational invariance of the field similar to \cref{eq:matter-PS}, and the Cartesian power spectrum only depends on $k$ and $\mu=k_{||}/k$. 

Decomposing the 3D Dirac-delta function in terms of $\bk$ into $\ell,\phi_\ell$, and $n$ with \cref{eq:3D-k-n,eq:DD-polar}, we have
\begin{align}
    \delta^{\rm D}(\bk-\bk')=\frac{\Delta x\, x_{\rm eff}^2}{\pi\ell}\delta^{\rm D}(n-n')\delta^{\rm D}(l-l')\delta^{\rm D}(\phi_l-\phi_l')\,,\nonumber
\end{align}
then
\begin{align}
    &C_{\ell n n'}\approx\frac{1}{2\pi}P(k,\mu,x_{\rm eff})\delta^{\rm D}(n-n')\delta^{\rm D}(\ell-\ell')\nonumber\\
    &\qquad\qquad i^{m-m'}\int_0^{2\pi} \dd\phi_l\,\;e^{-i(m-m')\phi_{\ell}} \,\nonumber\\
   &\approx P(k,\mu,x_{\rm eff})\delta^{\rm K}_{nn'}\delta^{\rm K}_{\ell\ell'}\delta^{\rm K}_{mm'}\,.
\end{align}
Therefore, the Cartesian power spectrum $P(k,\mu)$ approximates the diagonal components of the SFB PS, and we have justified our results in \cref{sec:numerical-approx}.

 \begin{figure*}[htbp]
\centerline{\includegraphics[width=0.81\hsize]{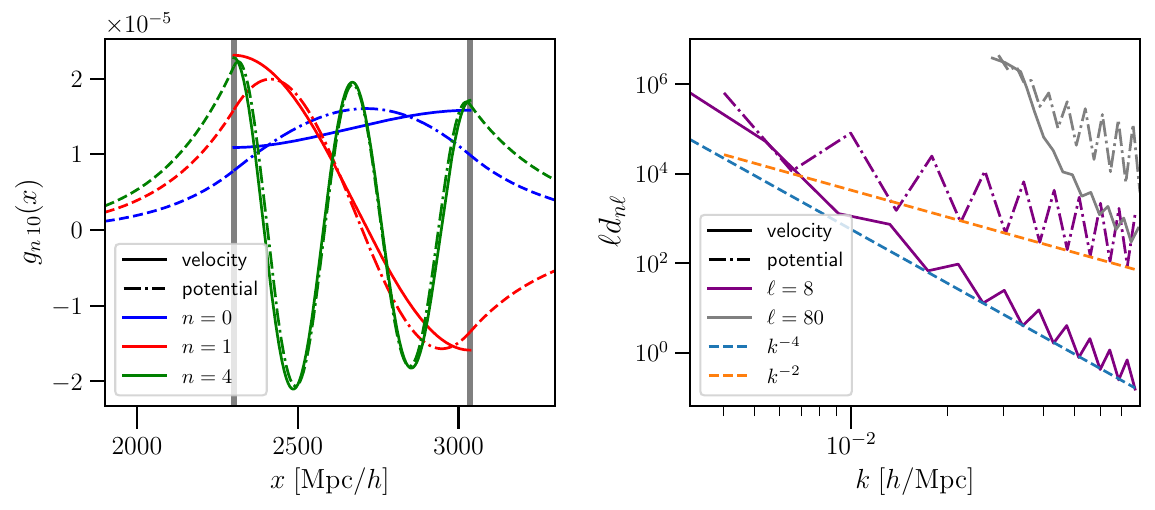}}
\caption{Comparison of SFB radial basis functions under the velocity and potential boundary conditions. The left panel shows $g_{n\ell}(x)$ at $\ell = 10$ for both boundary conditions. The vertical grey bands indicate the comoving radial range $[x_{\rm min}, x_{\rm max}]$ corresponding to $z = 1.0-1.5$, over which the basis functions are constructed. The solid lines represent basis functions satisfying the velocity boundary condition, under which the derivatives vanish at the boundaries. The dot-dashed lines correspond to the potential boundary condition, which enforces continuity with the solutions of the Laplace equation outside the boundary: $(r/r_{\rm min})^\ell$ for $r<r_{\rm min}$, and $(r_{\rm max}/r)^{\ell+1}$ for $r>r_{\rm max}$ \cite{21Gebhardt_SuperFab}. These external solutions are shown as dashed lines outside the boundary region. The right panel shows the scaling behavior of the radial coefficients $d_{n\ell}$ from \cref{eq:dnl} as a function of $k$, for fixed angular multipoles, under the two boundary conditions.}
\label{fig:potential_velocity_compare}
\end{figure*}

 \begin{figure}[htbp]
\centerline{\includegraphics[width=\hsize]{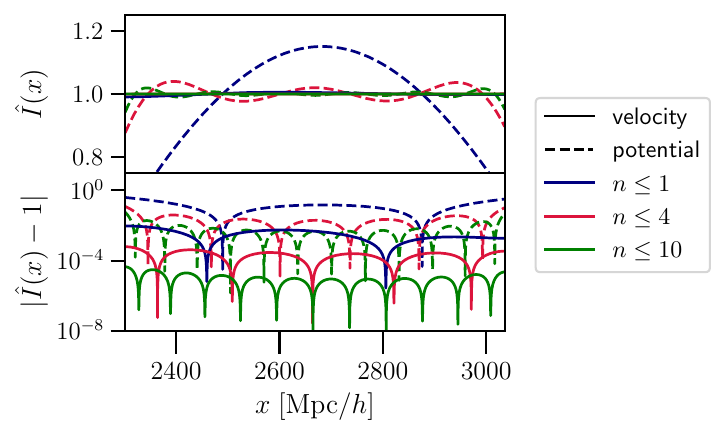}}
\caption{Reconstruction of the unit function $\hat{I}_N(x) = \sum_{n=0}^{N} d_{n\ell} \, g_{n\ell}(x)$ using SFB radial basis functions at $\ell = 10$, up to order $N$, under the velocity and potential boundary conditions. The upper panel shows the reconstructed functions, while the lower panel displays the absolute reconstruction error. As the order $N$ increases, both sets of basis functions converge toward the unit function. However, the velocity boundary condition yields significantly higher accuracy at each order and therefore achieves convergence more rapidly.}
\label{fig:potential_velocity_reconstruct}
\end{figure}

\section{Choice of Boundary Conditions}\label{sec:velocity-over-potential}

In this work, we have chosen the velocity (Neumann) boundary condition over the potential boundary condition. We here compare the two choices and discuss the benefit of velocity boundary condition in the power spectrum analysis. 

The potential boundary condition \cite{95Fisher_SFB,21Gebhardt_SuperFab} solves the Laplace's equation\footnote{Laplace’s equation is Poisson’s equation without a source term. The field $\delta(\bx)$ satisfies the Poisson’s equation inside the SFB volume.} outside the spherical shell and ensures the density contrast to be continuous and smooth on the boundaries (see the left panel of \cref{fig:potential_velocity_compare} for illustration). Therefore, the potential boundary condition is the preferred choice for density or velocity reconstruction \cite{95Fisher_SFB,24Blake_velocity_SFB} with the SFB basis, since the continuity and smoothness on the boundaries of the reconstructed fields are desired. However, such properties are not relevant for power spectrum analysis.  

The velocity boundary condition in \cref{eq:velocity-boundary-def} sets the derivative of the density field, that is the velocity, to vanish on the boundaries. This ensures a sphere in real space will remain a sphere in redshift space \cite{23Gebhard_SFB_eBOSS}, which is not true for the potential boundary condition. The radial basis functions exactly match the phase of cosine functions under the velocity boundary condition (see \cref{fig:gnl_zero,fig:gnl_pattern} along with \cref{eq:g0}), while the potential boundary condition introduces additional dips at both ends of the domain, as shown in the left panel of \cref{fig:potential_velocity_compare}, to satisfy continuity with Laplace equation solutions outside the boundaries. Despite both solving the Laplace eigenequations, we emphasize that the radial functions under the two boundary conditions are distinct radial bases and can yield different decompositions. Due to the better match with cosine functions, the velocity boundary condition yields more accurate plane-parallel approximation under \cref{eq:pp-SFB} compared to the potential boundary condition, so the SFB PS give a better estimate for the Cartesian clustering wedge. 

The two boundary conditions yield similar SFB power spectra for cosmological signals \cite{23Gebhard_SFB_eBOSS}, and both choices should yield consistent cosmological results as long as one adopts the same boundary condition for both estimator and theory. However, they can exhibit different behavior for systematics. For an angular systematic with uniform radial distributions, the diagonal components of the SFB PS for a fixed angular multipole fall as $k^{-8}$ under the velocity boundary condition as shown in \cref{fig:SFB_foreground_line}, due to the radial coefficients behaving as $d_{n\ell}\sim k^{-4}$. In contrast, the potential boundary condition yields $k^{-4}$ for the SFB PS, corresponding to $d_{n\ell}\sim k^{-2}$. The scaling behavior of these radial coefficients $d_{n\ell}$ (defined in \cref{eq:dnl} for the unit function) under the two boundary conditions is shown in the right panel of \cref{fig:potential_velocity_compare}.
 
To better understand this difference, we examine the reconstruction of the unit function $\hat{I}_N(x) = \sum_{n=0}^{N} d_{n\ell}\, g_{n\ell}(x)$ using a truncated basis of radial modes for a fixed angular multipole under both boundary conditions. As shown in \cref{fig:potential_velocity_reconstruct}, while both sets of basis functions eventually converge to the unit function as $N$ increases, showing both sets of basis functions contain all the information necessary to reconstruct the function in the real space if $N\to\infty$, the basis under the velocity boundary condition converges significantly faster and achieves higher accuracy at a given finite order. This better convergence reflects the faster decay of the corresponding radial coefficients $d_{n\ell}$ with respect to $n$, which in turn explains the stronger suppression of angular systematics in the smaller radial modes for the SFB PS under the velocity boundary condition.

In conclusion, purely angular systematics have less leakage to smaller radial modes and contaminate fewer modes in the SFB PS under the velocity boundary condition. We therefore recommend the use of the velocity boundary condition over the potential one, since angular systematics can be more localized in SFB space.

 \begin{figure*}[htbp]
\centerline{\includegraphics[width=0.9\hsize]{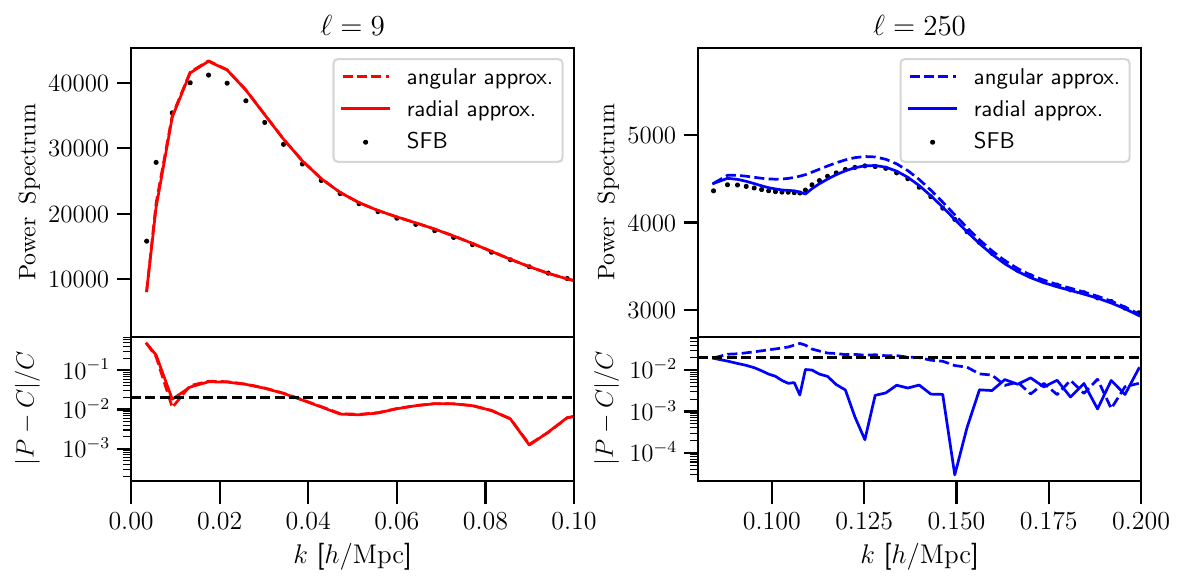}}
\caption{Comparison between the exact SFB PS and its plane-parallel approximation with $P(k,\mu)$ under the angular approximation based on $\ell$ (\cref{eq:pp-SFB-angular}) and the radial approximation based on $n$ (\cref{eq:pp-SFB}) for a redshift bin from 1.0 to 1.5. The bottom panel shows the absolute value of the relative difference between the two spectra, and the black dashed line indicate $2\%$ relative error. The radial approximation gives more accurate estimates for the SFB PS at small angular scales. The kink at around $k\approx 0.11 h/{\rm Mpc}$ in the right panel for the SFB PS at $\ell=250$ is due to the transitory behavior of the SFB radial basis functions discussed in \cref{fig:gnl_pattern}: the transition distance $x_{n\ell}^{\rm t}$ remains at $x_{\rm min}$ for $k\gtrapprox 0.11$.}
\label{fig:SFB_compare_Pkmu_schemes}
\end{figure*}

 \begin{figure}[htbp]
\centerline{\includegraphics[width=\hsize]{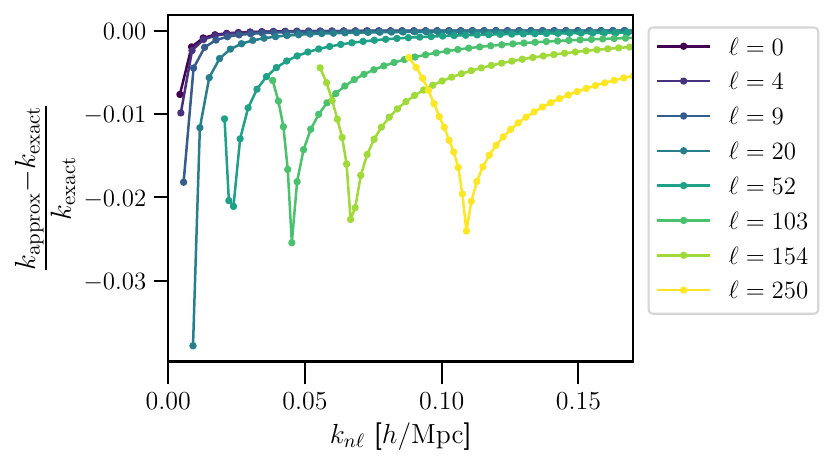}}
\caption{The approximation of $k_{n\ell}$ using the angular and modes $\ell$ and $n$ described by \cref{eq:k-approx} for a SFB volume with $z$ from 1.0 to 1.5. }
\label{fig:k_approx_ln}
\end{figure}

 \section{Approximating Transverse Modes}\label{sec:angular-approx}

In \cref{sec:radial}, we relied on the radial index $n$ to define the LOS Fourier mode $k_{||,n\ell}$ (\cref{eq:k-LOS}) and then subtract the LOS mode from the total Fourier mode to obtain the transverse Fourier mode $k_{\perp,n\ell}$ (\cref{eq:k-pp}). Here we explore the use of angular multipoles $\ell$ to directly define the transverse modes, and we have
\begin{align}
    k_{\perp,n\ell}^{\rm ang.}=\frac{\ell+\frac{1}{2}}{x_{{\rm eff},n\ell}}\,,
    \label{eq:k_transverse_l}
\end{align}
where $x_{{\rm eff},n\ell}$ is the effective distance for each SFB mode defined in \cref{eq:eff-xnl}, we use the ``ang." superscript to distinguish this approximation based on $\ell$ from the default approximation based on the radial index $n$, and we have replaced $\ell$ in \cref{eq:angular_approx} with $\ell+\frac{1}{2}$ to better match the eigenvalues of Laplacian\footnote{To match the eigenvalues of the angular Laplacian $\nabla^2_\theta$ and the 2D Cartesian Laplacian $\nabla^2_\perp$, we set $r_0^2 k_\perp^2 = \ell(\ell+1)$, which yields $k_\perp r_0 = \sqrt{\ell(\ell+1)} = \ell(1+1/\ell)^{1/2} \simeq \ell+1/2 + 
\mathcal{O}(1/\ell)$ \cite{20Gebhardt_non-linear}.}.

We can therefore use the above transverse Fourier mode defined through $\ell$ to approximate the SFB power spectrum:
\begin{align}
    C_{\ell nn}^{{\rm ang.}\,{\rm approx.}} \equiv P\left(k=k_{n\ell},\mu=\mu^{\rm ang.}_{n\ell},x_c=x_{{\rm eff},n\ell}\right)\,,
    \label{eq:pp-SFB-angular}
\end{align}
and the LOS $\mu$ factor now depends on $\ell$ through
\begin{align}
   \mu^{\rm ang.}_{n\ell}= \sqrt{1-\left(\frac{k_{\perp,n\ell}^{\rm ang.}}{k_{n\ell}}\right)^2}\,.
\end{align}
We refer to the above equations as the angular approximation (through $\ell$) to the SFB PS, an alternative to the radial approximation (through $n$) defined in \cref{eq:pp-SFB}. 

We have used the radial approximation as the default option in the main text, since it gives noticeably better approximations to the SFB PS at small angular scales, as we show in \cref{fig:SFB_compare_Pkmu_schemes}. In particular, the radial approximation captures the correct shape of the exact SFB PS at the large scales of the high angular multipoles (the right panel), while the angular approximation fails to do. The angular approximation does yield comparable results at large angular scales (the left panel). However, for the goal of plane-parallelization, the radial approximation in \cref{eq:pp-SFB} is the better option.

Using the transverse (\cref{eq:k_transverse_l}) and LOS Fourier modes (\cref{eq:k-LOS}) that depend on angular and radial indices respectively, we can approximate the magnitude of the total Fourier modes as
\begin{align}
k_{n\ell}^2&\approx (k_{\perp,n\ell}^{\rm ang.})^2+k_{||,n\ell}^2\nonumber\\
&=\left(\frac{\ell+\frac{1}{2}}{x_{{\rm eff},n\ell}}\right)^2+\left(\frac{n\pi}{x_{\rm max}-x_{n\ell}^{\rm t}}\right)^2\,.
\label{eq:k-approx}
\end{align}
We show the accuracy of the above approximation in \cref{fig:k_approx_ln}, and using $\ell$ and $n$ indices alone are able to characterize the total Fourier mode in each SFB to the percent level. However, there are still significant patterns in the residuals of the approximation, suggesting missing or misrepresented components in \cref{eq:k-approx}. 

We have explored using different definitions of effective distance $x_{\rm eff}$ for approximating the transverse Fourier modes in \cref{eq:k_transverse_l}, but they do not improve the accuracies for either the plane-parallelization of SFB PS or the approximation to the total Fourier modes, albeit with different residual patterns. For a relatively broad redshift bin like $z=1.0-1.5$, it is difficult to determine the exact location of effective distance that is required for converting angular modes $\ell$ to transverse Fourier modes $k_{\perp}$, causing the inaccuracies we observe in \cref{fig:SFB_compare_Pkmu_schemes,fig:k_approx_ln} at small angular scales. In comparison, the relationship between the LOS Fourier modes $k_{||}$ and the radial index $n$ is well-defined for a redshift bin of any size, which explains the better performance of the radial approximation for plane-parallelizing the SFB PS.

\section{Useful Idenitities}
The Jacobi-Anger expansion
\begin{align}
   e^{i{\bold l}\cdot\hat{\bn}}&=\sum_{m}i^m J_m(\ell\theta)e^{im(\phi-\phi_{\ell})}
   \label{eq:JA-expansion}
\end{align}
Mean of spherical harmonics
\be
\label{eq:Y_mean}
    \int \dd^2 \hat{\bx}~ Y_{\ell m} (\hat{\bx}) = \sqrt{4\pi}\delta_{\ell0}\delta_{m0}
\ee
Dirac-delta in polar coordinates
\be
\label{eq:DD-polar}
    \delta^{\rm D}(\br-\br')=\frac{1}{r}\delta^{\rm D}(r-r')\delta^{\rm D}(\phi-\phi')
\ee

\bibliography{refs}

\end{document}